\documentclass[twocolumn]{aastex631}
\usepackage{gensymb}
\usepackage{longtable}

\newcommand\target{HIP\,65426\,b}
\newcommand\rref{HIP\,68245}
\newcommand\lbol{$\mathrm{log}\!\left(L_\mathrm{bol}/L_{\odot}\right)$}
\newcommand\sklip{\texttt{spaceKLIP}}
\newcommand{\fref}{Figure~\ref}
\newcommand{\tref}{Table~\ref}
\newcommand\rev[1]{#1}%{\textcolor{purple}{\textbf{#1}}}

\shorttitle{\footnotesize{\textit{JWST}} \normalsize imaging of \footnotesize{HIP\,65426\,}\normalsize{\MakeLowercase{b}} from 2$-$16 micron}
\shortauthors{Carter et al.}
\graphicspath{{./}{figures/}}
\begin{document}

\title{The \textit{JWST} Early Release Science Program for Direct Observations of Exoplanetary Systems I: \\ High-Contrast Imaging of the Exoplanet HIP\,65426\,b from 2$-$16 \boldmath$\mu$m}

\author[0000-0001-5365-4815]{Aarynn L.~Carter    }\affiliation{Department of Astronomy \& Astrophysics, University of California, Santa Cruz, 1156 High St, Santa Cruz, CA 95064, USA}
\author[0000-0001-8074-2562]{Sasha Hinkley	  }\affiliation{University of Exeter, Astrophysics Group, Physics Building, Stocker Road, Exeter, EX4 4QL, UK}
\author[0000-0003-2769-0438]{Jens Kammerer	       }\affiliation{Space Telescope Science Institute, 3700 San Martin Drive, Baltimore, MD 21218, USA}
\author[0000-0001-6098-3924]{Andrew Skemer	       }\affiliation{Department of Astronomy \& Astrophysics, University of California, Santa Cruz, 1156 High St, Santa Cruz, CA 95064, USA}	
\author[0000-0003-4614-7035]{Beth A.~Biller      }\affiliation{Scottish Universities Physics Alliance, Institute for Astronomy, University of Edinburgh, Blackford Hill, Edinburgh EH9 3HJ, UK; Centre for Exoplanet Science, University of Edinburgh, Edinburgh EH9 3HJ, UK}

\author[0000-0002-0834-6140]{Jarron M.~Leisenring}\affiliation{Steward Observatory and the Department of Astronomy, The University of Arizona, 933 N Cherry Ave, Tucson, AZ, 85721, USA}
\author[0000-0001-6205-9233]{Maxwell A.~Millar-Blanchaer}\affiliation{Department of Physics, University of California, Santa Barbara, CA, 93106}
\author[0000-0003-0331-3654]{Simon Petrus	       }\affiliation{Instituto de F\'{i}sica y Astronom\'{i}a, Facultad de Ciencias, Universidad de Valpara\'{i}so, Av. Gran Breta\~{n}a 1111, Valpara\'{i}so, Chile} \affiliation{N\'{u}cleo Milenio Formac\'{i}on Planetaria - NPF, Universidad de Valpara\'{i}so, Av. Gran Breta\~{n}a 1111, Valpara\'{i}so, Chile}	
\author[0000-0003-0454-3718]{Jordan M.~Stone     }\affiliation{Naval Research Laboratory, Remote Sensing Division, 4555 Overlook Ave SW, Washington, DC 20375 USA}
\author[0000-0002-4479-8291]{Kimberly Ward-Duong }\affiliation{Department of Astronomy, Smith College, Northampton, MA, 01063, USA}	
% ...



\begin{abstract}
We present \textit{JWST} Early Release Science (ERS) coronagraphic observations of the super-Jupiter exoplanet, \target{}, with the Near-Infrared Camera (NIRCam) from 2$-$5~$\mu$m, and with the Mid-Infrared Instrument (MIRI) from 11$-$16~$\mu$m. At a separation of $\sim$0.82\arcsec~(86$^{+116}_{-31}$~au), \target{} is clearly detected in all seven of our observational filters, representing the first images of an exoplanet to be obtained by \textit{JWST}, and the first ever direct detection of an exoplanet beyond 5~$\mu$m. \rev{These observations demonstrate that \textit{JWST} is exceeding its nominal predicted performance by up to a factor of 10, depending on separation and subtraction method, with measured 5$\sigma$ contrast limits of $\sim$1$\times10^{-5}$ and $\sim$2$\times10^{-4}$ at 1\arcsec{} for NIRCam at 4.4~$\mu$m and MIRI at 11.3~$\mu$m, respectively}. These contrast limits provide sensitivity to sub-Jupiter companions with masses as low as 0.3$M_\mathrm{Jup}$ beyond separations of $\sim$100~au. Together with existing ground-based near-infrared data, the \textit{JWST} photometry are well fit by a \texttt{BT-SETTL} atmospheric model from 1$-$16~$\mu$m, and span $\sim$97\% of \target{}'s luminous range. Independent of the choice of model atmosphere we measure an empirical bolometric luminosity that is tightly constrained between \lbol=$-$4.31 to $-$4.14, which in turn provides a robust mass constraint of 7.1$\pm$1.2~$M_\mathrm{Jup}$. In totality, these observations confirm that \textit{JWST} presents a powerful and exciting opportunity to characterise the population of exoplanets amenable to high-contrast imaging in greater detail. 
\end{abstract}

\correspondingauthor{Aarynn L. Carter}
\email{aarynn.carter@ucsc.edu}


%%%%%%%%%%%%%%%%%%%%%%%%%%%%%%%%%%%%%%%%%%%%%%%%%%%%%%%%%%%%%%%%%%%%%%%%%%%%%%%%%%%%%%%%%%%%%%%%%%%%%%%%%%%%%%%%
% INTRODUCTION
%%%%%%%%%%%%%%%%%%%%%%%%%%%%%%%%%%%%%%%%%%%%%%%%%%%%%%%%%%%%%%%%%%%%%%%%%%%%%%%%%%%%%%%%%%%%%%%%%%%%%%%%%%%%%%%%
\section{Introduction}\label{sec:intro}
Across the last twenty-five years a variety of observational techniques have been employed to uncover and characterise the current population of over 5000 confirmed exoplanets \citep{ConfirmedPlanets}. Of these techniques, the direct detection of photons from an exoplanetary atmosphere -- direct imaging -- remains one of the most challenging due to the substantial contrast in flux between host stars and their exoplanetary companions. The emitted flux of an exoplanet can be many magnitudes fainter than the stellar diffraction halo at its angular separation, and bespoke instrumentation (e.g., \citealt{Maci14, Beuz19}) and/or image post-processing (e.g., \citealt{Chau04, Maro08}) is needed to isolate the exoplanet emission. Even with state-of-the-art instruments and data analysis procedures, only $\sim$20 planetary mass companions (PMCs) have been detected and characterised through these ``high-contrast'' observations \citep{Curr22}, and all exoplanets directly imaged to date have estimated or dynamically-measured masses $\gtrsim$2~$M_\mathrm{Jup}$.

Despite these drawbacks, high-contrast observations offer considerable advantages to other techniques. At present, high contrast imaging is the most viable technique for characterising the population of exoplanets at orbital separations greater than $\sim$10~au. Furthermore, beyond the large population of irradiated, close-in planets with atmospheric measurements obtained via exoplanet transit observations, direct observations of exoplanet emission remain the most readily accessible path towards the characterisation of exoplanet atmospheres. Constraints on atmospheric composition may improve our understanding of exoplanet formation and evolution (e.g., \citealt{Ober11, Morl19, Zhan21, Moll22}), although these determinations can be highly dependent on the post-formation accretion of solid material. Compared to close-in transiting exoplanets, directly imaged planets present a distinct advantage in this regard, as they are easier to detect at younger ages where they are less likely to have experienced significant migration and/or accretion. Additionally, at young ages bulk properties such as temperature, radius, and bolometric luminosity provide independent constraints on formation conditions \citep{Marl07, Marl14} that can be contrasted to atmosphere-driven conclusions on formation. Finally, the study of exoplanet atmospheres continues to advance towards smaller, and more Earth-like exoplanets, and could ultimately lead to the discovery of life outside our Solar System \citep{Schw18}. 

Nevertheless, to fully realise the advantages of high-contrast imaging, upgraded or new observatories and instruments (e.g., \citealt{Gard06, Male18, Chil20}) will be necessary so that we can expand this technique across a broader wavelength range, and to a wider diversity of closer separation and/or lower mass exoplanets.

%At present $\sim$20 planetary mass companions (PMCs) have been detected and characterised through high-contrast observations \citep{Curr22}, and all exoplanets directly imaged to date have estimated or dynamically-measured masses $\gtrsim$2~$M_\mathrm{Jup}$. Despite the small sample size, this subset of objects has driven significant developments in our overall understanding of exoplanet atmospheres and architectures. For example, the formation of exoplanets through gravitational instability is rare \citep{Viga17, Niel19, Viga21}; brown dwarfs and exoplanets may exhibit different eccentricity distributions \citep{Bowl20}; clouds and disequilibrium chemistry influence measured spectra (e.g., \citealt{Skem12, Apai13, Morl14}); similar to brown dwarfs, exoplanets can be variable \citep{Zhou19, Zhou20, Vos22}; wide separation exoplanets may be more prevalent within systems that also host a circumstellar disk \citep{Mesh17}; and transits of exomoons may be observable around isolated planetary mass objects \citep{Limb21}. Nevertheless, upgraded or new observatories and instruments (e.g., \citealt{Gard06, Male18, Chil20}) will be necessary to directly image and characterise a wider diversity of closer separation and/or lower mass exoplanets at a higher precision and across a broader wavelength range.

\subsection{High-Contrast Observations with \textit{JWST}}
%Launched on December 25th, 2021, \textit{JWST} \citep{Gard06} is an international collaboration between NASA, ESA, and the CSA, and the first large strategic mission of the NASA Astrophysics Division to launch since the 1990's \citep{Missions}. 

With a primary mirror diameter of 6.5~m, an operational wavelength range from 0.6$-$28.1~$\mu$m, and a diverse range of instrumental modes, \textit{JWST} \citep{Gard06} presents a revolutionary opportunity for scientific exploration and discovery across many branches of astronomy and astrophysics. Within this remit, high-contrast observations of exoplanets and exoplanetary systems are no exception.

\textit{JWST} is located at the second Sun$-$Earth Lagrange point, far from the thermal background, telluric contamination, and wavefront aberrations generated by Earth's atmosphere. The combination of excellent optical performance  ($\sim$75$-$130~nm RMS wavefront error, depending on the instrument), highly stable wavefront ($<$2~nm drift over a few hours), and large telescope aperture, enables \textit{JWST} to reach better photometric and spectroscopic limiting sensitivities than both past or current ground- (i.e., 8$-$10~m class telescopes) and space-based (e.g., \textit{Hubble}, \textit{Spitzer}) observatories \citep{Rigb22}. It is not only this increased sensitivity that improves our ability to detect and characterise faint objects such as exoplanets, but its combination with \textit{JWST}'s access to the near- and mid-infrared. At these wavelengths, the flux emitted from a hotter host star steadily decreases as a function of increasing wavelength, whereas the flux emitted from cooler exoplanetary companions reaches a peak. Hence, the natural star-planet contrast is minimised. To realise these advantages, \textit{JWST} offers a selection of instrumental modes designed for, or that can be applied to, high-contrast observations. Specifically, both NIRCam \citep{Riek05} and MIRI \citep{Riek15} are equipped with coronagraphic masks \citep{Kris09, Bocc15}, NIRISS \citep{Doyo12} is equipped with a non-redundant mask which enables aperture masking interferometry (AMI; \citealt{Siva12}), and although lacking any hardware for starlight suppression, both NIRSpec \citep{Bagn07} and MIRI are equipped with integral field units (IFUs; \citealt{Well15, Boke22}). 

In anticipation of the range of capabilities that \textit{JWST} would provide, a similar range of predictions and simulations were constructed in an effort to forecast its potential for exoplanet imaging science. With \textit{JWST}'s extraordinary sensitivity across 4$-$15$\mu$m (where cooler planets are more luminous, \citealt{Morl14}), the first direct detections of sub-Jupiter, and even sub-Saturn, mass planets will be within reach \citep{Bran20, Beic20, Cart21a, Ray23}. Already discovered companions will also be readily detectable across this broad wavelength range, allowing for deeper atmospheric characterisations that may result in the detection of a range of atmospheric species \citep{Dani18, Pata22}, or tighter constraints on bulk planetary properties. 

Nevertheless, these predictions are based on ground-based testing and observatory simulations, whereas the true capabilities of \textit{JWST} hinge on its on-sky performance. Preliminarily, the performance of both the NIRCam and MIRI coronagraphic modes exceeded expectations during observatory commissioning \citep{Gira22, Kamm22, Bocc22}, but the first scientific demonstrations of \textit{JWST}'s capabilities are being conducted as part of the Director's Discretionary Early Release Science (ERS) Programs\footnote{\href{https://www.stsci.edu/jwst/science-execution/approved-ers-programs}{https://www.stsci.edu/jwst/science-execution/approved-ers-programs}}.

%In anticipation of the range of capabilities that \textit{JWST} would provide, a similar range of predictions and simulations were constructed in an effort to forecast its potential for exoplanet imaging science. \citet{Bran20} predict that Saturn and Jupiter mass planets should be detectable with MIRI coronagraphy at 1$-$5~au separations across a sample of nearby ($\lesssim$10~pc) M-dwarfs. Similarly, \citet{Cart21a} demonstrate that both NIRCam and MIRI coronagraphy may be sensitive to sub-Jupiter mass planets for the majority of stars within the $\beta$~Pictoris moving group ($\beta$PMG) and TW Hydrae Association (TWA), albeit at separations $>$20$-$50~au. Ray et al. (submitted) further expands on the work of \citet{Cart21a}, and shows that sub-Jupiter mass planets could be detectable from 1$-$20~au for several stars within $\beta$PMG, TWA, and the Taurus Auriga Association with NIRISS AMI. At even closer separations, \citet{Beic20} predict that MIRI coronagraphy of $\alpha$~Centauri~A may be sensitive to $\sim$5~R$_\oplus$ companions from 0.5$-$2.5~au.  \citet{Dani18} demonstrate that already discovered companions \citep[e.g., HR~8799\,b/c/d][]{Maro08, Maro10}, $\beta$~Pictoris\,b \citep{Lagr10}) should also be detectable with MIRI coronagraphy, and NH$_3$ absorption could be identified for a subset of targets. Finally, \citet{Pata22} show that molecular mapping is possible with the MIRI IFU and may result in the detection of atmospheric species such as H$_2$O, CO, CO$_2$, CH$_4$, NH$_3$, and PH$_3$.

Our ERS program ``\textit{High Contrast Imaging of Exoplanets and Exoplanetary Systems with JWST}" (ERS-01386, \citealt{Hink22}) is the only ERS program that has tested the high-contrast exoplanet imaging modes of \textit{JWST} and includes: coronagraphic imaging from 2$-$16~$\mu$m of the known exoplanet \target{} (\citealt{Chau17}, this work) and circumstellar disk HD\,141569\,A (\citealt{Wein99}, Millar-Blanchaer et al. in preparation, Choquet et al. in preparation), spectroscopy from 1$-$28~$\mu$m of the PMC VHS\,J125601.92-125723.9\,AB\,b (VHS\,1256\,b, \citealt{Gauz15, Mile23}), and AMI observations of HIP\,65426 at 3.8~$\mu$m (Sallum et al. in preparation, Ray et al. in preparation). This program is rapidly disseminating these crucial initial data, and demonstrating the true capabilities of \textit{JWST} for high-contrast imaging and spectroscopy for the first time. Furthermore, we will provide a range of science enabling products (e.g., data analysis pipelines, recommendations for best observing practices) to the community to support their own proposals and investigations in Cycle 2 and beyond \citep{Hink22}. 

In this work we focus exclusively on the coronagraphic imaging observations of the HIP\,65426 system within this ERS program, and their context within a broader understanding of \textit{JWST} as a tool for high-contrast imaging.

\subsection{HIP\,65426\,b}
Discovered by \citet{Chau17}, \target{} is a super-Jupiter mass exoplanet at a wide physical separation of 110$^{+90}_{-30}$~au \citep{Chee19} to the star HIP\,65426 (A2V, 2MASS\,$J$=6.826, $J$$-$$K$=0.055, $M_\odot$=1.96$\pm$0.04). HIP\,65426 is located at a distance of 107.49$\pm$0.40~pc \citep{Gaia20}, has no signs of binarity from radial velocity and sparse aperture masking observations \citep{Chau17, Chee19}, and is a fast rotator ($v\mathrm{sin}(i)$=299$\pm$9~km\,$s^{-1}$, \citealt{Chau17}). Furthermore, HIP\,65426 is a likely member of the Lower Centaurus-Crux association as derived from its proper motion and radial velocity measurements (89\% probability, \citealt{Gagn18a}), constraining its age to 14$\pm$4~Myr. The interest in this association has grown over time with the increasing number of directly imaged exoplanet discoveries within it (e.g., HD\,95086\,b \citealt{Rame13}, PDS~70\,b/c \citealt{Kepp18}, and TYC~8998\,b/c \citealt{Bohn20}. 

\begin{figure}
    \centering
    \includegraphics[trim={0.65cm 3.55cm 0.5cm 0.5cm},clip,width=\columnwidth]{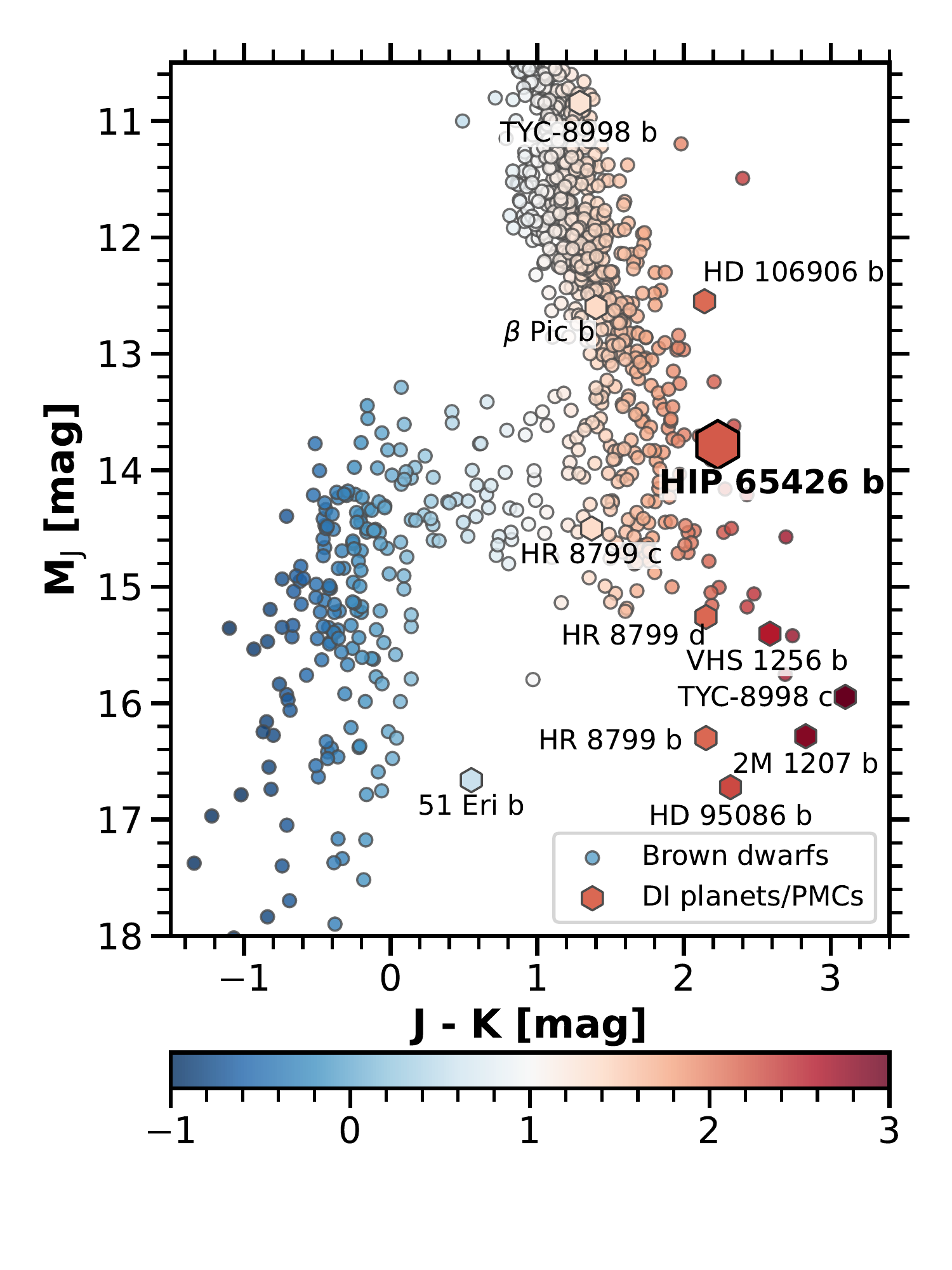}
    \caption{Colour-magnitude diagram showing the position of \target{} \citep{Chau17} relative to both the population of brown dwarf objects (circles) and a selection of directly imaged planetary mass companions (PMCs, hexagons), as obtained from \citet{Best20}.}
    \label{fig:cmd_jk}
\end{figure}

Although \target{} was initially observed with a combination of low-resolution spectroscopy and photometry from $\sim$1$-$2~$\mu$m \citep{Chau17}, follow up observations expanded this coverage to $\sim$1$-$5~$\mu$m \citep{Chee19,Stol20}, including a medium-resolution spectrum from $\sim$2$-$2.5~$\mu$m (R$\simeq$5500, \citealt{Petr21}). Photometric analysis has demonstrated that \target{} is similarly located to mid-to-late L-dwarfs in colour-magnitude diagrams (Figure \ref{fig:cmd_jk}), and lies between already discovered early L-dwarf exoplanet companions (e.g., $\beta$~Pic\,b, HD\,106906\,b) and those at the L/T transition (e.g., HR~8799\,c/d/e). Using combined photometric and spectroscopic observations, \citet{Petr21} performed an atmospheric forward model analysis of \target{}, indicating that it has $T_\mathrm{eff}$=1560$\pm$100~K, log($g$)$\leq$4.40~dex, [M/H]=0.05$^{+0.24}_{-0.22}$~dex, and the atmospheric carbon-to-oxygen ratio has an upper limit of C/O$\leq$0.55. Furthermore, \citet{Petr21} also detect the presence of H$_2$O and CO in the atmosphere of \target{} using cross-correlation molecular mapping, in addition to non-detections of CH$_4$ and NH$_3$. Finally, from evolutionary model analyses to the data reported in \citet{Chau17}, \citet{Marl19} estimate the mass of \target{} to be 9.9$^{+1.1}_{-1.8}$~$M_\mathrm{Jup}$ or 10.9$^{+1.4}_{-2.0}$~$M_\mathrm{Jup}$ for hot- or cold-start initial entropy conditions, respectively (see e.g., \citealt{Marl07, Spie12}). 

%Given its impressive sensitivity, \textit{JWST} offers a unique opportunity to characterise \target{} at a high precision across the near- to mid-infrared wavelength range. Prior to this work, observations beyond $\sim$5~$\mu$m have never been obtained for \target{}, or any other directly imaged exoplanet.

%Probing this wavelength range would provide valuable constraints on \target's broader emission profile, which may in turn signal the presence and extent of cloud absorption and flux redistribution in its atmosphere. 

%Although dependent on a variety of model assumptions, and with a low measurement precision, the determined C/O ratio for \target{} is indicative of formation via core accretion \citep{Ober11}, and may suggest that \target{} formed much closer to its host star (where the formation of an object this massive is predicted to be more likely) before migrating outwards. Nevertheless, it is still possible that it may have formed at wider separations through gravitational instability and then accreted a significant proportion of solid material \citep{Petr21}. 

%Despite the utility of these existing data, the limited precision of the observations beyond $\sim$3~$\mu$m has prohibited a more detailed analysis of the composition of \target's atmosphere. Between 3$-$5~$\mu$m in particular, prominent absorption features driven by CH$_4$, CO, and CO$_2$ may be detectable. More detailed constraints on these species would improve our knowledge on the chemical inventory for \target{}, constrain the presence and strength of disequilibrium chemistry effects, and strengthen the precision on the relative C/O abundance measurement \citep{Hink22}. Furthermore, 

\vspace{4pt}
In Section~\ref{sec:obs} we describe our \textit{JWST} observations of \target{} and all necessary data reduction. In Section~\ref{sec:analysis} we describe the analysis steps taken to produce residual starlight subtracted images and measurements of contrast performance. We present a discussion of these observations in the context of both the overall performance of \textit{JWST} in Section~\ref{sec:performance}, and our understanding of \target{} in Section~\ref{sec:hip65426b_results}. Finally, we summarise our conclusions in Section~\ref{sec:conclusion}.

%%%%%%%%%%%%%%%%%%%%%%%%%%%%%%%%%%%%%%%%%%%%%%%%%%%%%%%%%%%%%%%%%%%%%%%%%%%%%%%%%%%%%%%%%%%%%%%%%%%%%%%%%%%%%%%%
% OBSERVATIONS
%%%%%%%%%%%%%%%%%%%%%%%%%%%%%%%%%%%%%%%%%%%%%%%%%%%%%%%%%%%%%%%%%%%%%%%%%%%%%%%%%%%%%%%%%%%%%%%%%%%%%%%%%%%%%%%%
\section{Observations \& Data Reduction}\label{sec:obs}
The NIRCam and MIRI coronagraphic imaging observations of \target{} presented here were taken as part of program ERS-01386 \citep{Hink22} and exist as a subset of a broad range of observations to assess the performance of \textit{JWST}'s high-contrast imaging and spectroscopic modes with respect to the study of exoplanetary systems \citep{Hink22}.

\subsection{Observational Structure}\label{sec:obsstruct}
The observational strategies used for this program were adopted following the recommended best practices as known prior to launch and described in the \textit{JWST} user documentation\footnote{\href{https://jwst-docs.stsci.edu/jwst-near-infrared-camera/nircam-observing-strategies/nircam-coronagraphic-imaging-recommended-strategies}{https://jwst-docs.stsci.edu/jwst-near-infrared-camera/nircam-observing-strategies/nircam-coronagraphic-imaging-recommended-strategies}}$^,$\footnote{\href{https://jwst-docs.stsci.edu/jwst-mid-infrared-instrument/miri-observing-strategies/miri-coronagraphic-recommended-strategies}{https://jwst-docs.stsci.edu/jwst-mid-infrared-instrument/miri-observing-strategies/miri-coronagraphic-recommended-strategies}}. All observations of \target{} are repeated at two independent roll angles separated by $\sim$10\degree{} to enable subtraction of the residual stellar point spread function (PSF) through angular differential imaging \citep[ADI,][]{Mull85, Liu04, Maro06}. Although a large number of rolls across a larger angular range would be desirable for an optimal subtraction using this technique, the combination of lengthy exposure times, increased overheads, and spacecraft orientation constraints prohibit an observing strategy more complex than described. Given the maximum possible roll offset for \textit{JWST} at any given epoch is 14$\degree$, a larger roll offset than we have adopted would also require multi-epoch observations. 

We also perform similar observations of a bright reference star, HIP\,68245 (B2IV, 2MASS\,$J$=4.628, \mbox{$J$$-$$K$=0.137}), to additionally enable subtraction through reference differential imaging (RDI). This star was selected as it is: a) bright, therefore reducing the exposure time required to attain a similar signal-to-noise as the science target, b) a similar spectral type to the science target, therefore reducing the impact of spectral type mismatch, c) is relatively close ($\sim$10$\degree$) to the science target, therefore reducing slew overheads and minimising position dependent wavefront drift between science and reference observations, and d) has no evidence of binarity as determined by \textit{VLT}/SPHERE AMI observations (Proposal ID: 108.22CD). For information on selecting a suitable reference star, see the \textit{JWST} User Documentation\footnote{\href{https://jwst-docs.stsci.edu/methods-and-roadmaps/jwst-high-contrast-imaging/jwst-high-contrast-imaging-proposal-planning/hci-psf-reference-stars}{https://jwst-docs.stsci.edu/methods-and-roadmaps/jwst-high-contrast-imaging/jwst-high-contrast-imaging-proposal-planning/hci-psf-reference-stars}}. All exposure settings for the science and reference observations are shown in \tref{tab:observations}.

\begin{table*}
    \centering
    \begin{tabular*}{\textwidth}{@{\extracolsep{\stretch{.09}}}*{11}{c}@{}}
        \hline
        \hline
        \vspace{1mm}
        Filter & $\lambda_\mathrm{mean}$ ($\mu$m) & W$_\mathrm{eff}$ ($\mu$m) & Mask & Readout & $N_\mathrm{groups}$ & $N_\mathrm{ints}$ & $t_\mathrm{exp}$ ($s$) & $N_\mathrm{dithers}$ & $N_\mathrm{rolls}$ & $t_\mathrm{total}$ ($s$) \vspace{1mm} \\
        \hline \vspace{-4mm}
        \textbf{\textbf{HIP\,65426}} & & & & \\
        \\
        F250M & 2.523 & 0.179 & MASK335R & DEEP8 & 15 & 4 & 1235.892 & 1 & 2 & 2471.784\\
        F300M & 3.067 & 0.325 & MASK335R & DEEP8 & 15 & 4 & 1235.892 & 1 & 2 & 2471.784 \\
        F356W & 3.580 & 0.769 & MASK335R & DEEP8 & 15 & 2  & 617.946 & 1 & 2 & 1235.892 \\
        F410M & 4.084 & 0.436 & MASK335R & DEEP8 & 15 & 2 & 617.946 & 1 & 2 & 1235.892 \\
        F444W & 4.397 & 0.979 & MASK335R & DEEP8 & 15 & 2  & 617.946 & 1 & 2 & 1235.892\\
        F1140C & 11.307 & 0.608 & FQPM1140 & FASTR1 & 101 & 41 & 1002.102 & 1 & 2 & 2004.204 \\
        F1550C & 15.514 & 0.703 & FQPM1550 & FASTR1 & 250 & 60 & 3609.341 & 1 & 2 & 7218.682\vspace{1mm}\\
        \textbf{\rref} & & & & \\
        F250M & 2.523 & 0.179 & MASK335R & MEDIUM8 & 4 & 4 & 166.852 & 9 & 1 & 1501.669\\
        F300M & 3.067 & 0.325 & MASK335R & MEDIUM8 & 4 & 4 & 166.852 & 9 & 1 & 1501.669\\
        F356W & 3.580 & 0.769 & MASK335R & MEDIUM8 & 4 & 2 & 83.426 & 9 & 1 & 750.835\\
        F410M & 4.084 & 0.436 & MASK335R & MEDIUM8 & 4 & 2 & 166.852 & 9 & 1 & 750.835\\
        F444W & 4.397 & 0.979 & MASK335R & MEDIUM8 & 4 & 2 & 83.426 & 9 & 1 & 750.835\\
        F1140C & 11.307 & 0.608 & FQPM1140 & FASTR1 & 52 & 10 & 126.791 & 9 & 1 & 1141.116 \\
        F1550C & 15.514 & 0.703 & FQPM1550 & FASTR1 & 100 & 19 & 459.706 & 9 & 1 & 4137.356\vspace{1mm}\\
        \hline
    \end{tabular*}
    \caption{Target and reference exposure settings. Background observations were also performed for the MIRI F1140C and F1550C filters with parameters identical to two exposures of a single roll or dither of the target and reference observations, respectively (see Section \ref{sec:obsstruct}). Filter mean wavelengths ($\lambda_\mathrm{mean}$) and bandwidths (W$_\mathrm{eff}$) are taken from \sklip~(see Section \ref{sec:spaceklip}). See \href{https://jwst-docs.stsci.edu/understanding-exposure-times}{https://jwst-docs.stsci.edu/understanding-exposure-times} for further detail on \textit{JWST} exposure settings.}
    \label{tab:observations}
\end{table*}

An RDI based subtraction is likely to reach superior contrast limits to ADI from pre-launch predictions \citep{Lajo16, Perr18, Cart21b}, and is therefore also more representative of the optimal performance of \textit{JWST} coronagraphy (also see Section \ref{sec:ach_contrast}). The exposure settings for these reference observations were chosen to reach an approximately equivalent fraction of full well saturation per integration as the corresponding target observations. Additionally, each reference observation was repeated at nine separate dither positions following small-grid dither patterns 9-POINT-CIRCLE and 9-POINT-SMALL-GRID for the NIRCam and MIRI observations, respectively \citep{Soum14, Lajo16}. The goal of this strategy is to produce a small library of reference PSFs for each science exposure which captures different misalignments between the star and the center of the coronagraphic mask, and can in turn facilitate more advanced PSF subtraction techniques (e.g., KLIP, \citealt{Soum12}). Further discussion on the relative benefits between ADI and RDI subtraction strategies, or a combination of the two, with respect to these \textit{JWST} observations can be found in Section \ref{sec:ach_contrast}. 

For MIRI, we also add background observations to both our science and reference observations in both filters of a nearby ``empty'' region of sky (as identified in WISE images, \citealt{Wrig10}) separated $\sim$1.5\arcmin{} away the target star to measure the stray light ``glow stick'' that is inherent to MIRI coronagraphic observations \citep{Bocc22}. Specifically, this position corresponds to a right ascension and declination of $\alpha$=13$\degree$24\arcmin{}44.2915\arcsec{}, $\delta$=$-$51$\degree$29\arcmin{}31.54\arcsec{}, respectively \rev{(ICRS J2000 coordinates)}. To best match the science and reference observations, the exposure parameters for each background observation exactly match the parameters for a single roll/dither of the associated science or reference target. These observations were intended to be performed at two separate dither positions to identify astrophysical sources that might impact the background subtraction, however, due to a previously unresolved issue they were instead repeated at an identical pointing (Dean Hines, private communication). 

The NIRCam and MIRI observations were executed as two separate non-interruptible sequences, ensuring that observations between rolls, and also between science, reference, and background targets, are minimally separated in time. This reduces the extent to which the wavefront can vary across observations, due to variations in the telescope mirror alignment, the thermal evolution of the telescope, or both \citep{Perr18, Cart21b}. Changes in the wavefront will lead to variations in the residual PSF between exposures, hinder our ability to perform an optimal subtraction of these residual PSFs, and suppress the overall achievable contrast. 

\subsection{spaceKLIP}\label{sec:spaceklip}
For all observations we perform data reduction using the newly developed and publicly available \texttt{python}  package, \sklip\footnote{\href{https://github.com/kammerje/spaceKLIP}{https://github.com/kammerje/spaceKLIP}} \citep{Kamm22}. Briefly, \sklip{} takes a collection of data products from the \texttt{jwst} pipeline\footnote{\href{https://jwst-pipeline.readthedocs.io}{https://jwst-pipeline.readthedocs.io}\label{fn-jwst}}$^,$\footnote{All data were processed using pipeline version, \texttt{CAL\_VAR}=1.9.4, and calibration reference data, \texttt{CRDS\_CTX}=jwst\_1041.pmap} \citep{jwstpipe} as inputs, and generates PSF subtracted images, contrast curves, and measurements of companion photometry and astrometry. The majority of this functionality is provided by the underlying \texttt{pyKLIP} \citep{Wang15} package, with \sklip{} providing a user friendly interface, streamlined code execution, custom \textit{JWST} data reduction routines, and built-in plotting procedures. 

%Furthermore, \texttt{pyKLIP} requires Stage 2 \texttt{jwst} data products as inputs, whereas \sklip{} can additionally accept Stage 0 and Stage 1 products$^{\ref{fn-jwst}}$. These lower-level data products are processed to Stage 2 within \sklip{} itself using the \texttt{jwst} pipeline, however, a selection of parameters can also be set to customise its execution. 
%\vspace{5mm}
\subsection{NIRCam Coronagraphy}\label{sec:nircam_coro}
The NIRCam observational sequence was executed from 23:00 July 29th to 05:16 July 30th 2022 UTC, with exposures taken using the MASK335R round coronagraphic mask \citep{Kris10} in the F250M, F300M, F410M, F356W, and F444W filters \rev{for the reference star (PA$=$110.2\degree), then HIP~65426 (PA$=$110.0\degree), and then finally HIP~65426 at a second roll angle (PA$=$120.4\degree)}. This sequence structure significantly reduces overheads, as once the target acquisition has been performed it is not necessary to reacquire the target to switch the observational filter. 

We begin data reduction using the Stage 0 (*uncal.fits) files as generated by the \texttt{jwst} pipeline. These products are then processed to Stage 1 (*rateints.fits) files using \sklip{}, which follows a slightly modified version of the \texttt{jwst} pipeline. Where possible the \textit{JWST} detectors have reference pixels that can be used to track and correct drifts in the measured pixel counts due to readout electronics. In the absence of such a reference, these drifts may instead be misinterpreted as ``jumps''\footnote{\href{https://jwst-pipeline.readthedocs.io/en/latest/jwst/jump/description.html}{https://jwst-pipeline.readthedocs.io/en/latest/jwst/jump/\\description.html}} from cosmic ray events during the up-the-ramp (MULTIACCUM) detector readout. The NIRCam coronagraphic subarrays do not have any embedded reference pixels and default pipeline processing leads to multiple erroneous jump detections and greatly increased noise in processed images. Therefore, we manually define all pixels within a four pixel border of each image as reference pixels within the pipeline to mitigate these effects. Additionally, we identify a significant improvement in image quality by skipping the dark current subtraction step that is turned on as a default in the \texttt{jwst} pipeline. \rev{At present, the dark current calibration data exhibit a large number of hot pixels, persistence, and cosmic rays which cannot be averaged out or corrected due to limited number of available integrations. Attempts to perform a dark current subtraction using this data result in a variety of negative flux residuals in the reduced images and an overall reduced sensitivity of the final pipeline product.} Once this calibration file is improved through further calibration observations, we do not anticipate a need to skip this step for NIRCam coronagraphic data.

During the jump detection step, the \texttt{jwst} pipeline will make use of a detection threshold value based on the estimated signal and noise to assess whether a deviation between groups is significant enough to be considered a jump (see~$^{\ref{fn-jwst}}$ for further detail). The default value for this threshold is 4, but we repeat an early version of our F444W analysis across thresholds of 4 to 16 to search for potential improvements. We find that the contrast is slightly improved from a threshold of 4 to 5 by $\sim$5$\times10^{-8}$ at $1$\arcsec, but at larger thresholds it does not vary (deviations $<$1$\times10^{-8}$ at $1$\arcsec). Hence, we adopt a detection threshold of 5 for all of our NIRCam analyses.

The Stage 1 products are processed further to Stage 2 (*calints.fits) files using \sklip{}, with some additional pixel cleaning procedures as follows. Firstly, every pixel with a data quality flag (e.g., indicating hot or warm pixels, unreliable data processing) is replaced by the median of its orthogonal and diagonal neighbours, with the notable exclusion of pixels with a jump flag which are typically grouped in clusters. We also inspect each pixel for temporal flux variations and identify situations where: a) the pixel is bright (MJy/sr $>$ 1) for at least one integration, and b) the pixel is relatively faint (a $>$80\% decrease in flux compared to the brightest integration) for at least one integration. The pixel values for the integrations that are not marked as faint are then replaced by the median value of the integrations that are marked as faint. We note here that an outlier identification process based on variations from the standard deviation of each pixel in time may be preferable, but is difficult to incorporate given the small number of integrations in these exposures (see \tref{tab:observations}). Despite the above corrections, $\sim$25 static hot pixels remain across all images. Although these pixels do not impact our ability to recover \target{}, they introduce residuals in the PSF subtraction process (see Section \ref{sec:psfsub}) that bias our measurements of the contrast performance. Therefore, we provide the locations of these pixels to \sklip{} manually and correct them in an identical manner as the pixels marked with data quality flags. 

As the NIRCam PSF in the F250M filter is undersampled, ringing artifacts are generated by the interpolation methods in \texttt{pyKLIP} used for image registration and spatial shifting of input PSFs. These artifacts bias the image registration process, and limit our ability to accurately inject synthetic PSFs for contrast curve calibration and companion astrometry and photometry. To overcome this issue, we smooth all of the F250M images by a Gaussian filter, as implemented by \texttt{scipy}, with $\sigma$=1.3~pixels. This value of $\sigma$ was chosen as it was the minimum possible value that removed the observed artifacts across a test of ten equally spaced values from $\sigma$=1 to $\sigma$=2. We note that this factor of 1.3 is consistent with the ratio of the detector pixel scale and the theoretical Nyquist sampling at 2.5~$\mu$m (assuming a reduced primary mirror diameter of 5.2~m due to the NIRCam Lyot stop). Smoothing these images may lead to reduced precision in our astrometric analysis, however, it should not influence the accuracy of our retrieved photometry.

\subsection{MIRI Coronagraphy}\label{sec:miricoro}
The MIRI observational sequence was executed from 21:05 July 17th 2022 to 05:19 July 18th 2022 UTC. \rev{Exposures were taken for HIP~65426 in the F1140C filter (PA$=$117.4\degree), then once again at a second roll angle (PA$=$108.0\degree), and then for the reference star (PA$=$109.2\degree). This sequence was then repeated in reverse for the F1550C filter, except with slightly different target roll angles of 108.2\degree and 117.4\degree.} This structure is different to that of the NIRCam observations as each filter is tied to a specific four-quadrant phase mask coronagraph (FQPM, \citealt{Bocc15}), and target acquisition must be repeated when switching between them. Inserting the reference observations between the observations of \target{} minimises the time separation between science and reference exposures for each filter, and therefore the extent of the wavefront evolution between them. After all science and reference observations are complete, we perform the dedicated background observations that are used to subtract the dominant ``glow stick'' stray light feature \citep{Bocc22} as described in Section \ref{sec:obsstruct}.

We begin data reduction using the Stage 0 (*uncal.fits) files as generated by the \texttt{jwst} pipeline. These products are then processed to Stage 1 (*rateints.fits) files using \sklip{}. Similarly to NIRCam (see Section \ref{sec:nircam_coro}) we explore the impact of the jump detection threshold on our analysis. For these data in particular, we found that the default jump detection threshold value of 4 is too low and leads to a number of pixels being erroneously flagged as containing a jump. Flagged pixels are interpreted differently to unflagged pixels in the ramp fitting procedure and the resulting Stage 1 files contain a large number of pixels with unrealistic (negative) flux values as a result. After repeating an early version of our F1140C analysis across thresholds of 4 to 16, we observe an improvement in contrast between a threshold value of 4 and 5 of $\sim$2$\times10^{-5}$ at 1\arcsec, and $\sim$1$\times10^{-5}$ at 2\arcsec. Beyond this value there is only slight improvement, and the obtained contrast varies by less than $\sim$5$\times10^{-6}$ at 1\arcsec. For our final analyses we select a threshold value of 8, as it has the best contrast at 2\arcsec{} and fewer pixels with unrealistic flux values than lower thresholds (as determined from visual inspection). 

\rev{Following ramp fitting we found that the first integration of each exposure contained a significantly increased level of non-uniform noise indicative of a detector reset charge decay anomaly, which is driven by differences in how the MIRI field effect transistors reset the detector charge prior to an exposure versus between integrations\footnote{\href{https://jwst-pipeline.readthedocs.io/en/latest/jwst/rscd/description.html}{https://jwst-pipeline.readthedocs.io/en/latest/jwst/rscd/\\description.html}}. Ideally this anomaly can be corrected by calibration dark exposures, however, the currently available calibration darks were acquired in quick succession, and do not have a similar amount of dead time (e.g., due to telescope slews/dithers) as our science exposures. As a result, during our observations the detector electronics were given a much longer time to settle, and our integrations exhibit an entirely different reset anomaly. This effect is most dominant for the first integration of each exposure, but appears to persist throughout the entire exposure as well. Quantitatively, the median flux of the first integration following background subtraction (see below) for the target/reference observations are 6.2/3.0$\sigma$ and 7.3/3.9$\sigma$ deviant from the average median flux across all integrations for the F1140C and F1550C observations, respectively. In comparison, the median flux of all other target/reference integrations deviates within ranges of 0.003$-$0.59/0.26$-$0.55$\sigma$ and 0.004$-$1.29/0.06$-$0.86$\sigma$ for the F1140C and F1550C observations, respectively. As this increased noise is not accurately captured in the available calibration files and differs significantly from all other integrations, we opt to exclude the first integration of each exposure from all further analysis. This exclusion corresponds to a 2.5\%/10\% and 1.7\%/5.3\% cut of the target/reference data for the F1140C and F1550C observations, respectively. In the future better dark exposures will be taken as part of observatory calibrations that may well remove the need to exclude the first integration, and we encourage future observers to carefully evaluate the calibration status of their data before adopting similar cuts.}

The Stage 1 products are processed further to Stage 2 (*calints.fits) files using \sklip{}, with some additional pixel cleaning procedures as follows. First, every pixel with a data quality flag (e.g., indicating hot or warm pixels, unreliable data processing) is replaced by the median of its orthogonal and diagonal neighbours, with the notable exclusion of pixels with a jump flag which are typically grouped in clusters. Following this correction, $\sim$30 static hot pixels remain in our images and are corrected following an identical procedure by manually providing the pixel locations to \sklip{}. Similarly to NIRCam, these final pixels primarily impact the measured contrast performance, and not our ability to recover \target{}.

\begin{figure}
    \centering
    \includegraphics[width=\columnwidth]{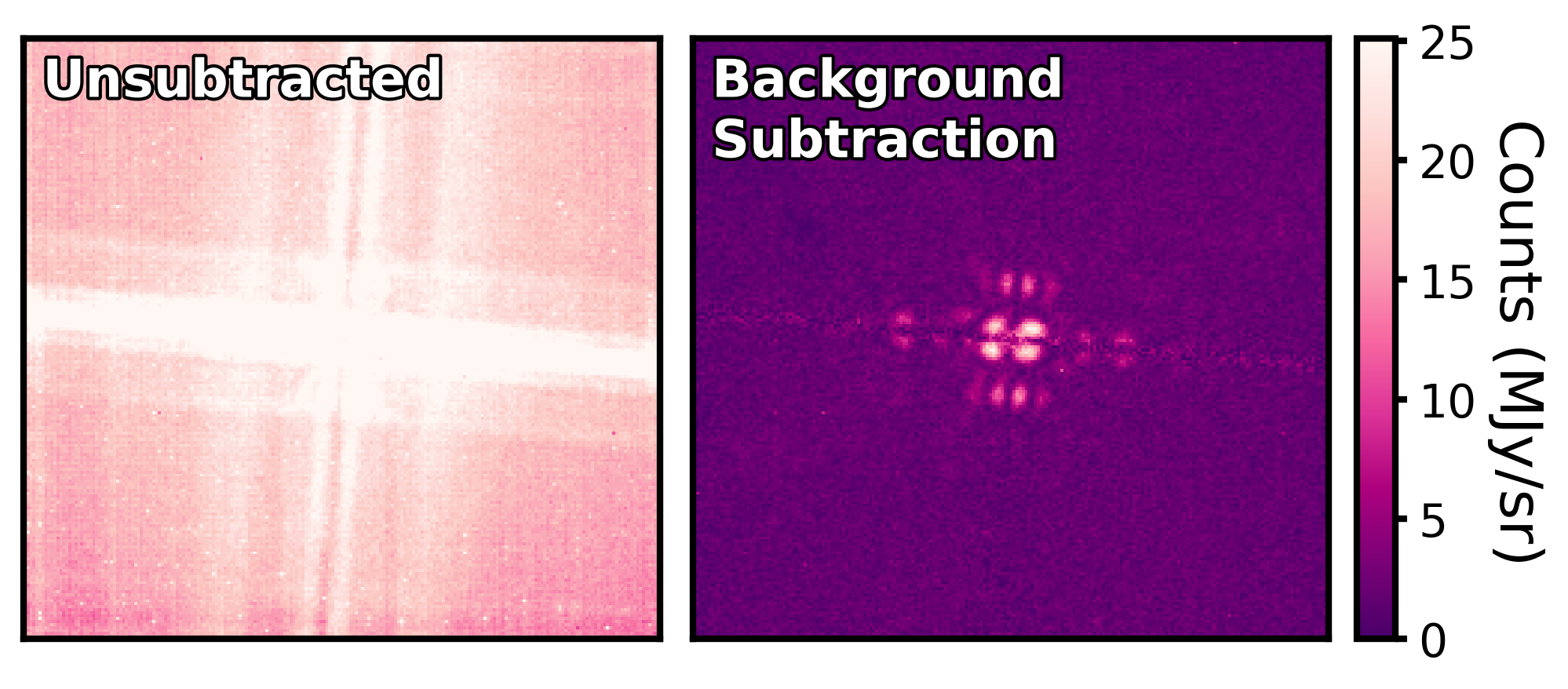}
    \caption{\textbf{\textit{Left:}} A single integration in the Stage 2 (*calints.fits) file for MIRI coronagraphy of HIP\,68245 in the F1550C filter. \textbf{\textit{Right:}} As on the left, except following subtraction of a median background frame of an ``empty'' region of the sky. Both images are identically scaled. Before subtraction the residual stellar flux is completely obscured by the stray light ``glow~stick'' \citep{Bocc22}, but can be easily recovered.}
    \label{fig:glowstick}
\end{figure}

As shown in \citet{Bocc22}, the MIRI coronagraphic fields of view are subjected to a stray light ``glow stick'' feature along the horizontal edges of the FQPMs that dwarfs the residual stellar flux (see \fref{fig:glowstick}). We subtract this feature from our processed Stage 2 products for each filter using a median background image of every 4$-$5 integrations from the dedicated background observations to the corresponding 4$-$5 science or reference integrations. The value of 4$-$5 was selected as it provided a slightly improved contrast at 1\arcsec{} compared to other tested numbers of integrations per median, ranging from 1 ($\sim$1$\times10^{-4}$ improvement) to all available integrations ($\sim$1$\times10^{-5}$ improvement). Grouping the median subtraction in this manner better captures the diffuse reset anomaly noise between integrations mentioned above. Additionally, it may help capture variations in the stray light feature, which varies with a standard deviation of $\sim$0.5\% (as identified by variations in the median pixel flux for pixels above 50\% of the peak pixel flux across integrations). 

Following the median background subtraction we find that the residual stellar flux is easily recovered (see \fref{fig:glowstick}). We also attempted to model the stray light glow stick as a principal component within both KLIP \citep{Soum14} and LOCI \citep{Lafr07} based subtractions, but found that they were susceptible to oversubtraction of the residual stellar flux and/or could not additionally account for background variations between integrations. We anticipate that with careful masking and optimisation of these algorithms it may be possible to overcome these issues, however, the median frame background subtraction is already highly effective and improvements to the achieved contrast are likely to be minimal. 

\subsection{Image Alignment}\label{sec:alignment}
The NIRCam and MIRI coronagraphic modes adopt independent target acquisition procedures to correctly center a star behind each focal plane mask. The in-flight positions of the NIRCam mask centers are known to better than $\sim$10~mas but the distortion model is still being refined. At present the target acquisition error for the MASK335R is as large as $\sim$12$-$30~mas, or 0.2$-$0.5~pixels \citep{Gira22}. This error is dominated by the precision of the centering algorithm, which is not well adapted to the PSF shape with coronagraphic optics (wider in $x-$axis), and not by the small angle maneuver (SAM) that places a target at its desired position behind the mask (which for NIRCam is repeatable to $\sim$6~mas). For MIRI the mask center positions have been measured to $\sim$5$-$10~mas, or $\sim$0.1~pixels \citep{Bocc22}. However, the SAM has a typical uncertainty of $\sim10-20$~mas, leading to positional offsets between different rolls or targets. Finally, between integrations the pointing stability of \textit{JWST} ($\sim$1~mas) and the accuracy of the small-grid dither manoeuvres ($\sim$2$-$4~mas) will lead to further positional shifts for both the NIRCam and MIRI coronagraphs \citep{Rigb22, Bocc22}.

%The NIRCam and MIRI coronagraphic modes adopt independent target acquisition procedures to correctly center a star behind the coronagraphic mask. The in-flight positions of the NIRCam mask centers are still being refined, and at present can have an error as large as $\sim$63$-$126~mas, or 1$-$2~pixels \citep{Gira22}. However, for MIRI the mask center positions have been measured to $\sim$5$-$10~mas, or $\sim$0.1~pixels \citep{Bocc22}. Moreover, the small angle maneuver that places a target at its desired position behind the coronagraphic mask has an typical uncertainty of $\sim10--20$~mas, leading to positional offsets between different rolls or targets. Finally, between integrations the pointing stability of \textit{JWST} ($\sim$1~mas) and the accuracy of the small-grid dither maneuvers ($\sim$2$-$4~mas) will lead to further positional shifts. By measuring these shifts, the absolute star position in each image can be determined, ensuring any measured astrometry is not biased by the centering error. Additionally, using these shifts we can realign images to a common center, which is important for maximising the efficacy of the NIRCam PSF subtraction. All image registration is performed on the integration level images. 

For NIRCam the absolute star position is only explicitly measured for the first science image in each filter. This position is measured using a cross correlation of a model coronagraphic PSF as obtained from \texttt{webbpsf\_ext} to the science PSF using the \texttt{scikit-image} package \citep{Walt14}. To best match the science PSF, the model PSF is generated using the telescope optical path difference (OPD) map as obtained on July 29th 2022. To identify the accuracy of this process we repeat the procedure, except comparing the model PSF to itself, and to a second model PSF that was generated using a pre-launch measurement of the telescope OPD (which exhibits comparable differences to the model PSF as our data). In each case, we manually shift the comparison PSFs across a range of 0.01 to 0.5 pixels and attempt to recover these offsets using the cross correlation process. For the self comparison, we recover the injected shift to at least $\sim$0.03~pixels, or 1.9~mas, whereas for the second model PSF comparison, we recover the injected shift to $\sim$0.1~pixels, or 6.3~mas. Given the comparable difference between the model PSF for both the second model PSF and our data, we adopt the latter uncertainty as a systematic uncertainty in our astrometric measurements (see \ref{sec:astrometry}). The shifts of the other science and reference images are obtained relative to the first science image through a similar cross correlation procedure. However, we instead use a box of 11$\times$11 pixels around the coronagraph center position, where the flux is dominated by the central core of the coronagraphic PSF. All measured relative shifts match expectations for the \textit{JWST} pointing precision of $\sim$1~mas (1$\sigma$, radial) \citep{Rigb22}.

\begin{figure*}[t]
    \centering
    \includegraphics[width=\textwidth]{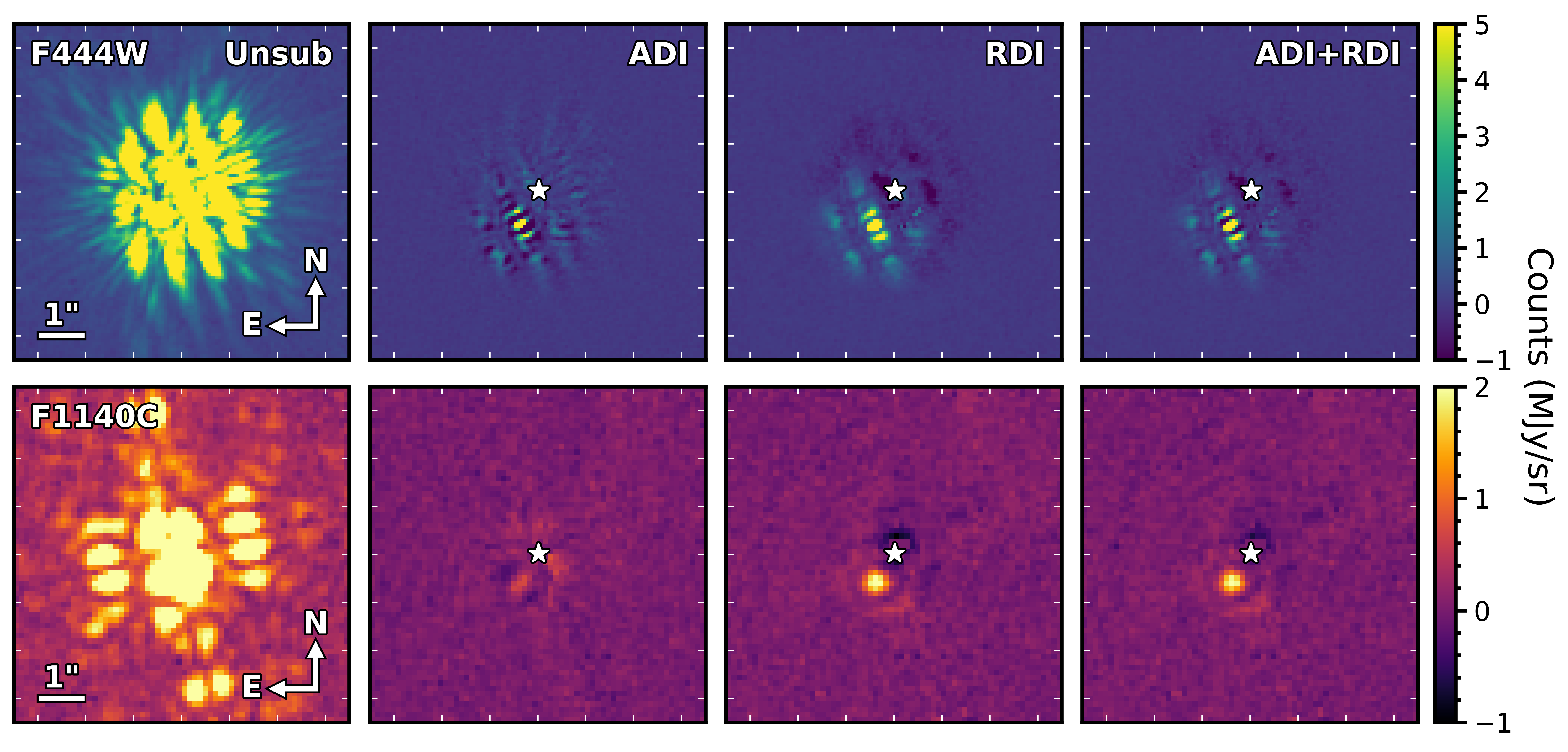}
    \caption{Unsubtracted and KLIP subtracted image stamps for the NIRCam \rev{F444W} (top row) and MIRI F1140C (bottom row) filters. The leftmost column displays the median unsubtracted image for a single science roll, and all other columns display the KLIP subtracted images for ADI, RDI, and ADI+RDI subtraction methods using the maximum number of KLIP PCA modes. All images are oriented as shown by the directional arrow in unsubtracted image column, the position of the star (white star) is also marked. Additionally, the intensity of all images for a given filter are identically scaled. The exoplanet, \target{}, can be easily identified at a position angle of $\sim$150\degree{} in the subtracted images. We note that the distinct ``hamburger'' shaped central core and six-lobed structure of the companion PSF in the NIRCam images is an expected feature that is related to the Lyot stop design, and not indicative of discrete astrophysical sources.}
    \label{fig:two_images}
\end{figure*}

To estimate the absolute star position for MIRI, we use the current measurements for the centers of the coronagraphic masks and assume the star is perfectly centered behind the coronagraphic mask. In zero-indexed subarray $x$-$y$ coordinates these values are (119.749, 112.236) and (119.746, 113.289) for the FQPM1140 and FQPM1550, respectively (Jonathan Aguilar, private communication). As a result, the stellar position is not known better than a minimum of $\sim$10~mas \citep{Bocc22}. Similarly, the relative alignment between images may be discrepant by $\sim$1$-$4~mas \citep{Rigb22}, or $<$0.05~pixels. Attempts to estimate the absolute star position in a similar manner to the NIRCam data were unsuccessful, and this process led to significantly larger estimated shifts than the known pointing stability of \textit{JWST} ($\sim$50~mas, vs $\sim$1~mas). This is most likely due to the variable spatial structure of the residual PSF, which significantly changes with small pointing shifts. An effort was also made to fit each MIRI image with model PSFs across all feasible pointings using \texttt{webbpsf\_ext}\footnote{\href{https://github.com/JarronL/webbpsf_ext}{https://github.com/JarronL/webbpsf\_ext}} within an \texttt{emcee}~\citep{Fore13} Markov chain Monte Carlo (MCMC) framework. However, these fits were unable to converge and upon visual inspection the spatial structure of the empirical and model PSFs were different, despite modelling the PSFs based on measurements of the telescope OPD within $\sim$1~day of our observations. We do not believe that using the coronagraphic mask centers as a proxy for the absolute star position has significantly affected our results, but it is certainly an area of improvement for future studies using MIRI coronagraphy. 

For NIRCam all images are aligned to a common center based on the measured shifts, however, for MIRI we opt to not perform any realignment. This decision is made under the assumption that because a pointing shift does not primarily cause a translation of the residual PSF in the MIRI images (in contrast to NIRCam), the unshifted reference images are more descriptive of variations in the science images. When comparing a RDI subtraction using unshifted reference images, and a separate RDI subtraction with a realignment based on the ideal small-grid dither positions, the measured contrasts are in agreement and this choice does not significantly impact our results. 

\subsection{PSF Subtraction}\label{sec:psfsub}
We perform a subtraction of the residual stellar PSF in each filter following three different principal component analysis (PCA) based methods as implemented in \sklip{}. First, we take the two independent rolls of \target{} and perform an ADI subtraction. Second, we perform an RDI subtraction by using the corresponding observations of the reference star, \rref, as a PSF library. Finally, we perform an ADI+RDI subtraction, which is identical to the RDI subtraction except that images at the opposite roll angle are also included in the PSF library. In each case the subtraction is performed on each integration from both science rolls individually, before being rotated to a common orientation as marked in \fref{fig:two_images} and summed together. Although the number of annuli and subsections the PSF subtraction is performed across can be adjusted, we find that this does not improve the observed contrast. Hence, we perform all subtractions using a single annulus and a single subsection (i.e., the entire image). \rev{We leave future optimisation of these parameters for future analysis, but note that any improvements to the measured contrast are likely to be small as the noise in our images is close to azimuthally symmetric.} The number of KLIP PCA modes can also be adjusted to tune the aggressiveness of the PSF subtraction. Hence, we perform the PSF subtraction across the full range of possible PCA modes to investigate the impact on our measured contrast and companion fitting. The maximum number of PCA modes is dependent on the exposure settings for each filter and corresponds to: the number of integrations in a single roll for ADI, the total number of integrations across all 9 dithers for RDI, and the sum of the two for ADI+RDI (see \tref{tab:observations} for precise values). 

Pre- and post-subtraction images for the NIRCam \rev{F444W} and MIRI F1140C filters are shown in \fref{fig:two_images}, and images for all filters are shown in Appendix \ref{app:images}. We note that the distinct ``hamburger'' shaped central core and six-lobed structure of the companion PSF in the NIRCam images is an expected feature that is related to the Lyot stop design, and not indicative of discrete astrophysical sources. 

%%%%%%%%%%%%%%%%%%%%%%%%%%%%%%%%%%%%%%%%%%%%%%%%%%%%%%%%%%%%%%%%%%%%%%%%%%%%%%%%%%%%%%%%%%%%%%%%%%%%%%%%%%%%%%%%%%%%%%%%%%%%%%%%%%%%%%%%%
% ANALYSIS / DISCUSSION %
%%%%%%%%%%%%%%%%%%%%%%%%%%%%%%%%%%%%%%%%%%%%%%%%%%%%%%%%%%%%%%%%%%%%%%%%%%%%%%%%%%%%%%%%%%%%%%%%%%%%%%%%%%%%%%%%%%%%%%%%%%%%%%%%%%%%%%%%%
\section{Analysis}\label{sec:analysis}

\subsection{Contrast Calibration}\label{sec:contrast_cal}
\begin{figure}
    \centering
    \includegraphics[width=\columnwidth]{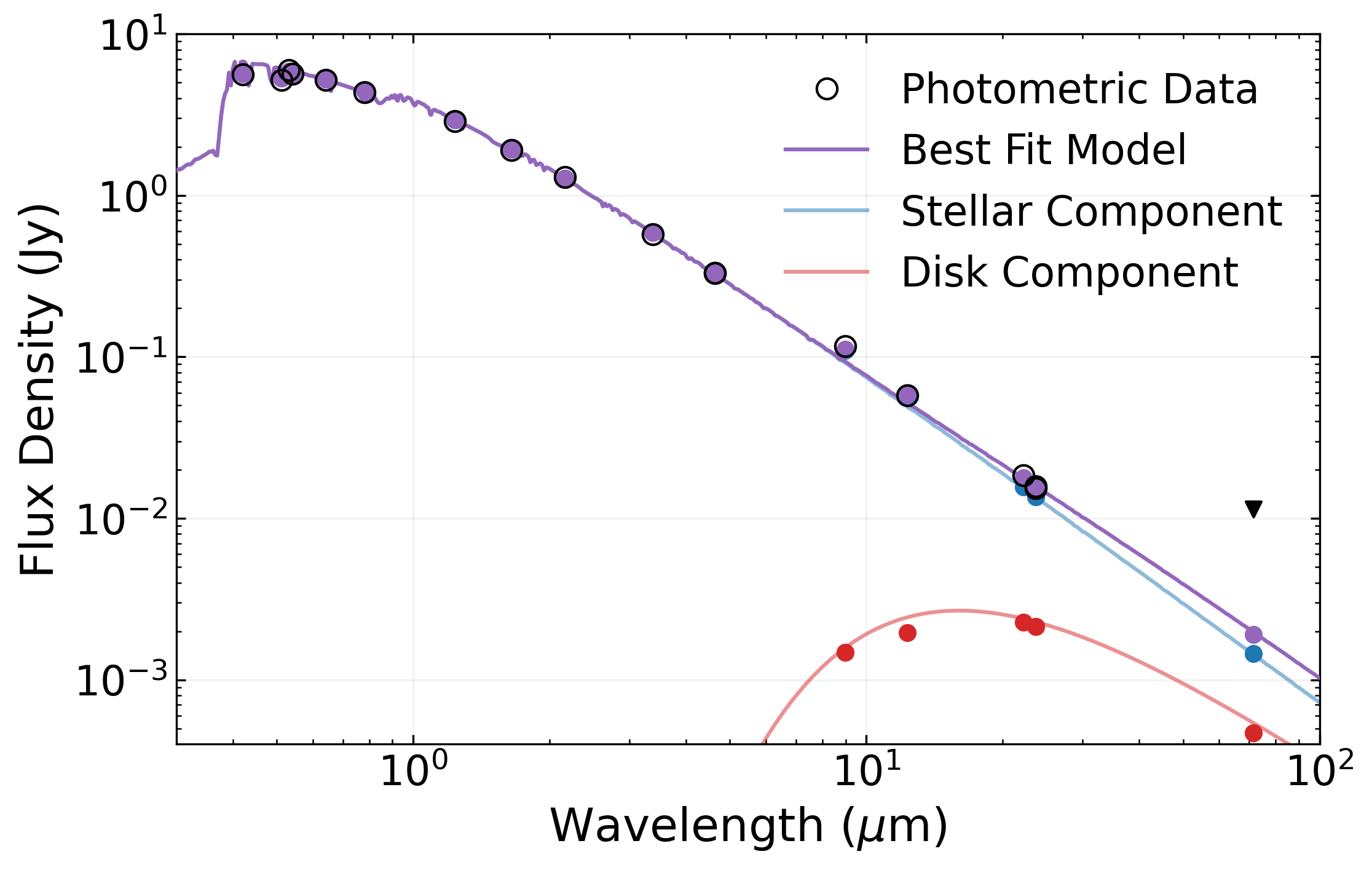}
    \caption{The best fit stellar model (purple) to existing: \textit{Hipparcos/Tycho-2}, \textit{Gaia} DR2, 2MASS, ALLWISE, \textit{AKARI} IRC, and \textit{Spitzer} MIPS data for HIP\,65426 (black circles). The stellar (blue) and disk (red) components of the model are also shown, along with the equivalent model fluxes in each photometric band (solid circles). Error bars are smaller than the circle diameter, except for the 70~$\mu$m MIPS point, which is instead marked as an upper limit (black triangle).}
    \label{fig:starspec}
\end{figure}

All proceeding contrast measurements are determined relative to a synthetic spectrum of HIP\,65426 in each of the \textit{JWST} filters, as estimated from fitting stellar and disk models to existing photometry following \citet{Yelv19}, see \fref{fig:starspec}. We use data of HIP\,65426 from \textit{Hipparcos/Tycho-2} \citep{Hog00}, \textit{Gaia} DR2 \citep{Gaia18}, 2MASS \citep{Cutr03}, ALLWISE \citep{Wrig10}, \textit{AKARI} IRC \citep{Ishi10}, and \textit{Spitzer} MIPS \citep{Chen12}. The fitting procedure compares synthetic photometry of models to the data to compute a $\chi^2$ value, and posterior distributions are found using \texttt{MultiNest} \citep{Fero09, Buch14}. We derive our own zero points using the CALSPEC Vega spectrum \citep{Bohl14}. We use PHOENIX models \citep{Alla12} for the stellar photosphere, and a Planck function for the disk model. There is a small excess at 24~$\mu$m that was previously reported at 3.5$\sigma$ by \citet{Chen12}, though not considered significant in that paper. The best fit model has an effective temperature of 8600$\pm$200~K and luminosity 16$\pm$1~$L_\odot$. The dust temperature and luminosity are very poorly constrained ($T_{\rm dust} = 300^{+200}_{-100}$~K, and $L_{\rm dust}/L_\star \sim 2 \times 10^{-5}$), though this uncertainty does not significantly impact the flux estimation in \textit{JWST} bandpasses because the excess is small. The flux excess at 24~$\mu$m is not high, but if real the dust would reside relatively close to the star, probably less than 
1~au (hence we use the total model flux to compute the contrast, as any dust component that contributes IR flux would remain unresolved). We use the posterior distribution of model parameters and synthetic photometry to generate a distribution of fluxes in the \textit{JWST} bands, and adopt the maximum likelihood solution for the stellar flux in each bandpass. 

\subsection{Contrast Curves}\label{sec:contrast}
Following PSF subtraction, we obtain metrics of the sensitivity as a function of angular separation (i.e., contrast curves) for all observational filters using \sklip. To avoid biasing the contrast measurement we mask regions of the subtracted images near \target{}, background sources, and the FQPM edges. These ``5$\sigma$" contrast curves report the flux level corresponding to a 5$\sigma$-equivalent false alarm probability of $2.9 \times 10^{-7}$ after correcting for small sample statistics at small separations \citep{Mawe14}. We call them 5$\sigma$ contrast curves for brevity.

To obtain a more accurate measurement of the contrast performance, the throughput of the coronagraph and the intrinsic throughput of the KLIP subtraction must be accounted for by injecting and then recovering the flux of artificial sources \citep{Adam20}. All artificial sources are generated using \texttt{webbpsf}\footnote{\href{https://webbpsf.readthedocs.io/en/latest/}{https://webbpsf.readthedocs.io/en/latest/}} \citep{Perr12,Perr14} at an initial intensity equivalent to a signal-to-noise ratio of 25. Immediately prior to injection and based on a desired injection location, each source is modulated by the coronagraphic throughput using a synthetic throughput map. 

Synthetic coronagraphic throughput maps are provided in the calibration reference files\footnote{\href{https://jwst-crds.stsci.edu}{https://jwst-crds.stsci.edu}}, however, both the provided NIRCam and MIRI FQPM maps are inaccurate. For the NIRCam MASK335R, the position of the occulting mask within the throughput map does not correspond with its actual location. Therefore, we modify the throughput map by extracting the pixels impacted by the occulting mask and repositioning them at the true mask center location of (149.9, 174.4) in zero-indexed subarray $x$-$y$ coordinates (Jarron Leisenring, private communication). \rev{In the case of MIRI, all of the FQPM maps are rudimentary and do not accurately capture the spatial throughput variations. Therefore, we instead use custom simulated maps of the FQPM throughput produced using \texttt{WebbPSF}. In brief, for each coronagraphic mode we generate 1681 position-dependent PSFs across a 25.6'' by 25.6'' grid spanning the subarray field of view (FOV). The PSFs are generated with logarithmic spacing such that they are most densely sampled along the FQPM axes. For each position in the FOV, we calculate two PSFs, one occulted and one unocculted, and take the ratio of the integrated PSF fluxes in each case to provide a throughput estimate at each position. From this 2D sampling of throughput, we use \texttt{scipy.interpolate.griddata} to linearly interpolate the throughput estimates across the FOV to produce a smooth 2D map matching the subarray dimensions for each MIRI FQPM.}

%In the case of MIRI, all of the FQPM maps are rudimentary and do not accurately capture the spatial throughput variations. Therefore, we instead use custom simulated maps of the FQPM throughput produced using \texttt{webbpsf\_ext}. In brief, \texttt{webbpsf\_ext} generates position-dependent PSFs based upon interpolation over a densely-sampled grid of individual PSFs along the FQPM axes. A uniform (flat) illumination of $10^{-7}$~Jy was convolved with the position-dependent PSFs to produce an illumination pattern that was then normalised and scaled to the subarray dimensions for each MIRI FQPM. 

\rev{Due to the target acquisition errors described in Section 2.5, the measured star centers are offset from the the center of the coronagraphic occulter. However, in addition to these pointing errors, for NIRCam coronagraphy we expect wavelength-dependent spatial shifts of the entire image as viewed on the detector focal plane due to refraction through the coronagraphic mask sapphire substrate along with deflections through filter optics further downstream in the optical train. Therefore, the star position in an image is not in isolation a suitable proxy for its position behind the coronagraphic mask. The wavelength-dependent shifts must be accounted for to more accurately determine the impact of the coronagraphic throughput across the detector focal plane and apply the correct throughput scaling to injected PSFs.} 

\rev{The aforementioned location for the MASK335R occulter at the detector focal plane is set based on the F335M filter, which is used for target acquisition and astrometric confirmation observations, and its image shift is defined to be zero. To measure the relative shifts of the remaining observational filters, we first determine the center of the projected stellar image for each filter observation of the science target roll positions and the reference target by cross-correlating an observed image in the Stage 1 *rate.fits file with a perfectly centered synthetic PSF generated with  \texttt{webbpsf}. These synthetic PSFs are created using the on-sky OPD map from July 29, 2022, and were recentered to remove pre-flight model shifts, which do not fully capture contributions from all optical elements, such as the different filter optics. The measured sub-pixel locations for each observed filter are then compared to the similarly measured F335M astrometric confirmation image to determine the filter-dependent offsets. The two independent target rolls and reference exposures provide three independent measures that are averaged together and presented in Table \ref{tab:filtoffs}; uncertainties are on the order of 1-2 mas. Finally, to apply the correct coronagraphic throughput to an injected synthetic planet in a given filter at a given position, we simply realign the throughput map according to the measured offset for that filter.}

Once scaled by the coronagraphic throughput, PSFs are injected into multiple copies of the unsubtracted science images across a range of separations extending to 4$\arcsec$, and for a range of position angles from 0 to 360$\degree$. Sources are not injected within 2$\lambda/D$ of each other or a masked region. These images then undergo KLIP subtraction in an identical manner to the science images, and the relative flux of an initial source PSF and the KLIP processed PSF as estimated within \texttt{pyKLIP} describes the overall coronagraphic mask plus KLIP throughput at each location. Finally, the basic 5$\sigma$ contrast is divided by an interpolation of the median throughput across all position angles to obtain the calibrated contrast. 

\begin{table}
    \centering\setlength{\tabcolsep}{18pt}
    \begin{tabular}{ccc}
        \hline
        \hline
        \vspace{1mm}
        Filter & $dx$ (pixels)  & $dy$ (pixels) \\
        \hline \vspace{-4mm} 
        \\
        F250M & 0.086$\pm$0.014 & -2.049$\pm$0.012 \\
        F300M & 0.078$\pm$0.014 & -0.531$\pm$0.016 \\
        F356W & 0.751$\pm$0.010 & -0.121$\pm$0.009 \\
        F410M & 0.177$\pm$0.013 & -0.086$\pm$0.021 \\
        F444W & 0.157$\pm$0.015 & -0.224$\pm$0.039 \\
        \hline
    \end{tabular}
    \caption{Filter dependent PSF offsets relative to the center of the NIRCam MASK335R as defined by the F335M filter.}
    \label{tab:filtoffs}
\end{table}

For the ADI subtraction, the measured contrast does not vary significantly when using more than 1 PCA mode for the NIRCam filters and more than 2 PCA modes for the MIRI filters. For RDI, the measured contrast does not vary beyond $\sim$6 modes for both NIRCam and MIRI. Finally, for ADI+RDI we find that there is a transition to improved contrast at $P_\mathrm{max}-P_\mathrm{ADI}$ modes, where $P_\mathrm{max}$ is the maximum possible number of PCA modes possible, and $P_\mathrm{ADI}$ is the maximum number of modes for the ADI subtraction. This is likely a result of the much larger number of reference images weighting the calculation of the PCA modes to be mostly RDI-like, until a sufficient limit is reached where the influence of the opposing roll images appears in the principal components. Beyond this transition value, the measured contrast does not vary significantly. 

Example calibrated contrast curves for the NIRCam \rev{F444W} and MIRI F1140C filters using the maximum number of PCA modes are shown in \fref{fig:contrast_curves}, and contrast curves for all filters are shown in Appendix \ref{app:ccurves}. 

\begin{figure*}
    \centering
    \includegraphics[width=\textwidth]{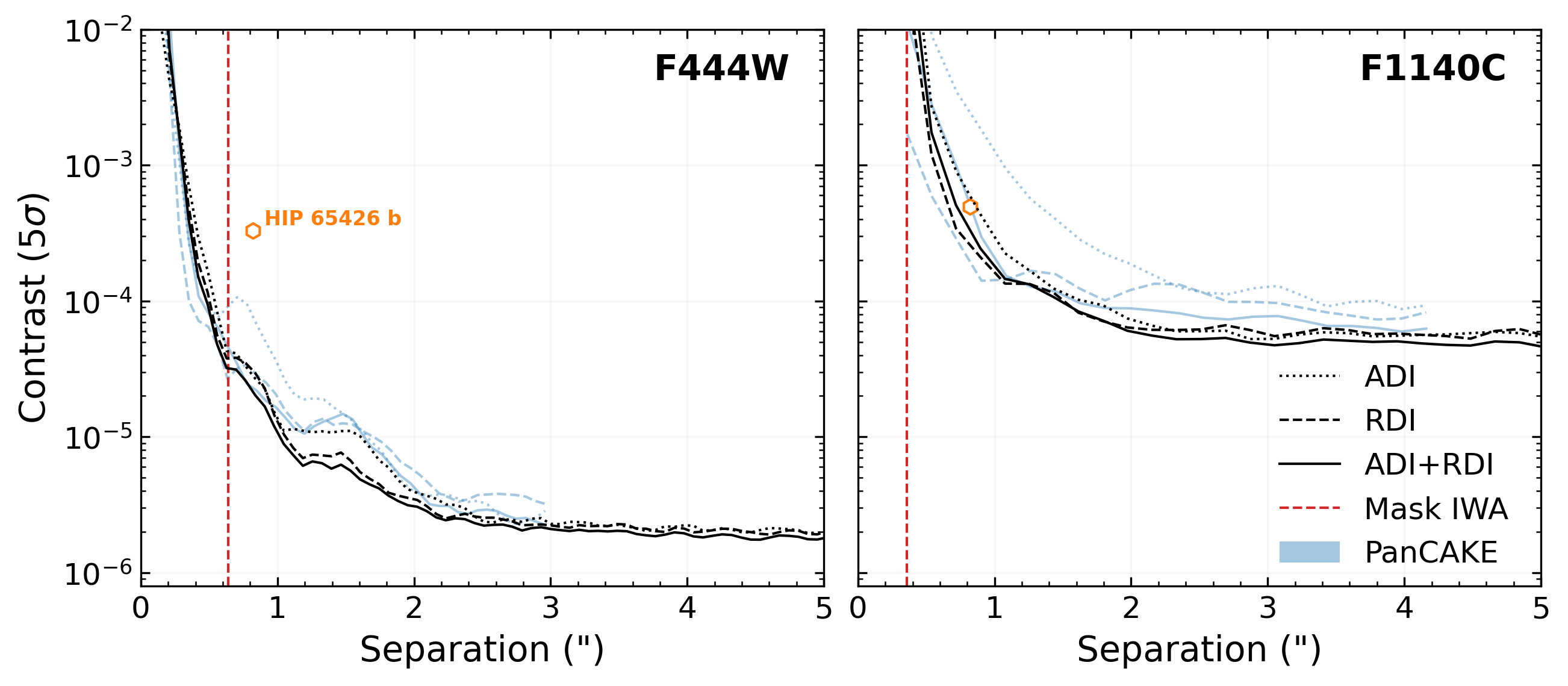}
    \caption{Contrast curves for observations in the \rev{F444W} and F1140C filters using both an ADI (dotted lines), RDI (dashed lines) and ADI+RDI (solid lines) subtraction using 20 PCA modes. Both the measured contrast of the true on-sky observations (black lines) and predicted contrasts as generated from PanCAKE \citep{Cart21b} (light blue lines) are displayed. The effective inner working angles (IWA) corresponding to the separation at 50\% transmission (red dashed lines), \rev{and the companion \target{} (orange hexagon)}, are also marked. Contrast curves for all other filters are displayed in \fref{fig:all_contrasts}.}
    \label{fig:contrast_curves}
\end{figure*}

\subsection{\target{} PSF Fitting}\label{sec:psffit}
To analyse the properties of \target{} in greater detail, we make use of the forward model PSF fitting routine provided by \texttt{pyKLIP} and implemented in \sklip. Briefly, this routine takes a model of the companion PSF and uses a forward model of the KLIP subtraction to apply PSF distortions that arise naturally from the KLIP process. This resultant PSF can then be scaled/shifted to best match the observed companion PSF and obtain a measurement of its location and intensity \citep{Puey16,Wang16}. 

For our analysis we adopt an independent model PSF for each filter using \texttt{webbpsf\_ext} functionality as implemented in \sklip. Specifically, we use \texttt{webbpsf\_ext} to generate an offset PSF at the predicted location of \target{} as adopted from \texttt{whereistheplanet} \citep{Wang21} and at the appropriate position for each science roll image. Each NIRCam and MIRI PSF is generated using a measured OPD map as determined from wavefront sensing and control observations on July 29th 2022 and July 17th 2022, respectively. Particularly for MIRI, the PSF of an off-axis source is still sensitive to its location relative to the FQPM edges, and explicitly generating a model PSF close to this location is important for obtaining a close match to the true companion PSF. As the spatial intensity of the model PSF depends on an assumed spectral energy distribution (SED), we use an existing best fit \texttt{BT-SETTL} model for \target{} from \citet{Chee19}. We note that as the initial model PSF is normalised to unit intensity, the primary purpose of selecting a model SED is not to accurately predict the absolute flux of \target{} in each bandpass, but instead to capture the relative variation in flux as a function of wavelength across each bandpass (which is largest for the broader NIRCam wide filters). \rev{Finally, the impact of the coronagraphic throughput on the received flux is applied by multiplying the normalised PSF by a scale factor equal to the relative difference between the integrated flux of the model PSF prior to normalisation and a matching \texttt{webbpsf\_ext} model PSF which excludes the coronagraphic elements.} 

The input model PSF is converted to physical units by scaling it to the flux of HIP\,65426 as estimated by the stellar model described in Section \ref{sec:contrast_cal}. Therefore, all derived photometry is anchored relative to our assumption of the stellar flux. Furthermore, any comparisons between the intensity of this PSF and the observed PSF is subject to an implicit assumption that the absolute flux calibration of \textit{JWST} is perfect. In reality, the absolute flux calibration accuracy requirement for NIRCam and MIRI coronagraphy is 5\% and 15\%, respectively\footnote{\href{https://jwst-docs.stsci.edu/jwst-data-calibration-considerations}{https://jwst-docs.stsci.edu/jwst-data-calibration-considerations}}. Both these fundamental uncertainties on the absolute flux calibration and the uncertainty on the stellar model flux as derived from the distributions described in Section \ref{sec:contrast_cal} are propagated as an increased uncertainty on all of our flux measurements. 

For each filter we fit the location and intensity of the model PSF to the true PSF of \target{} from the ADI+RDI subtraction using 6~PCA~modes. The fitting procedure is executed in an MCMC framework as implemented by \texttt{emcee} \citep{Fore13}, with 50 walkers for 2200 steps, of which the first 200 steps are discarded as burn-in. \sklip{} allows for the spatial scale of noise in an image to be fit with a variety of different kernels as implemented in a Gaussian process framework. For NIRCam the noise in our images appears to be correlated at the separation of \target{}, so, following \citet{Wang16}, we initially adopt a Mat\'ern 3/2 kernel to better capture this spatially correlated noise structure. However, the generated model PSF is slightly mismatched to the observed data and a positive flux residual was present at the PSF core. Future improvements to model PSF generation may alleviate this issue, but in this work we instead assume uncorrelated noise (using the ``diagonal kernel" option), which is able to better capture the flux of the companion at the expense of underestimating the obtained error bars. Therefore, more realistic error bars for the NIRCam photometric measurements are determined through a process of companion injection and recovery. For each filter, the best fit model PSF is used to subtract away the companion flux, and is then injected at 20 different position angles spanning 0$-$105\degree{} and 195$-$360\degree{} across an equivalent number of duplicate science images. The \target{} PSF fitting procedure is then repeated on these synthetic PSFs and the standard deviation in the measured flux across all 20 position angles is adopted as the estimated error. In the case of MIRI the observed noise is visually consistent with uncorrelated noise, so we adopt a diagonal kernel here as well. 

Results from the PSF fitting procedure are discussed further in Section \ref{sec:hip65426b_results}, and images of the data, model, and residuals to each PSF fit are displayed in Appendix \ref{app:psffits}.

%%%%%%%%%%%%%%%%%%%%%%%%%%%%%%%%%%%%%%%%%%%%%%%%%%%%%%%%%%%%%%%%%%%%%%%%%%%%%%%%%%%%%%%%%%%%%%%%%%%%%%%%%%%%%%%%%%%%%%%%%%%%%%%%%%%%%%%%%
% DISCUSSION %
%%%%%%%%%%%%%%%%%%%%%%%%%%%%%%%%%%%%%%%%%%%%%%%%%%%%%%%%%%%%%%%%%%%%%%%%%%%%%%%%%%%%%%%%%%%%%%%%%%%%%%%%%%%%%%%%%%%%%%%%%%%%%%%%%%%%%%%%%
\section{Instrument Performance}\label{sec:performance}

\subsection{Achieved Contrast}\label{sec:ach_contrast}
These observations provide a first look at the achieved contrast for \textit{JWST}'s high-contrast imaging modes following the completion of observatory commissioning and demonstrate that, in addition to the existing work of \citet{Gira22}, \citet{Kamm22}, and \citet{Bocc22}, both NIRCam and MIRI coronagraphic imaging are exceeding their predicted contrast performance. Examples of this improvement in comparison to pre-launch contrast estimates as obtained by \texttt{PanCAKE} \citep{Cart21b} are demonstrated in \fref{fig:contrast_curves}, and contrast curves for all seven filters are displayed in Appendix \ref{app:ccurves}.

At the MASK335R nominal IWA of 0.63\arcsec{}, the \rev{F444W data achieve a contrast of $\sim$4$\times10^{-5}$, sloping down to $\sim$1$\times10^{-5}$ at 1\arcsec{} and then $\sim$2$\times10^{-6}$ beyond 3\arcsec{}}. In comparison, at the FQPM1140C nominal IWA of 0.36\arcsec{}, the MIRI F1140C data achieve a contrast $\sim$1$\times10^{-2}$, sloping down to $\sim$2$\times10^{-4}$ at 1\arcsec{} and then $\sim$5$\times10^{-5}$ beyond 3\arcsec{}. For brevity we do not describe the achieved contrast in the other filters, and instead refer the reader to Appendix \ref{app:ccurves}.

In the background limited regime beyond 2\arcsec{}, the measured contrasts for NIRCam approximately match the predicted sensitivity of the ADI and ADI+RDI subtractions, and are up to 2 times more sensitive than the prediction for the RDI subtraction. For MIRI in this regime, all three subtraction methods outperform their predicted sensitivity by a factor of $1.5-2$. In the contrast limited regime below $\sim$2\arcsec{} we observe further improvements upon the predicted contrast, with the ADI and RDI subtractions demonstrating a factor of up to 5$-$10 times deeper contrast, and the ADI+RDI subtraction improving by a factor of $\sim$2. In some instances at the shortest separations below $\sim$0.6\arcsec{} the contrast does underperform by up to a factor of $\sim$2 compared to predictions for RDI subtractions. The primary driver for these improvements is likely the improvement in the overall optical and stability performance of \textit{JWST} compared to expectations. The total throughput is $\sim$10$-$20\% larger than predictions, driving analogous improvements in signal-to-noise; the overall telescope wavefront error is $\sim50$\% smaller than requirements (75/110~nm vs 150/200~nm for NIRCam/MIRI), improving the raw contrast; and the pointing stability of $1$~mas is $\sim$6$-$7~times smaller than predictions, meaning smaller drifts in alignment behind the coronagraphic mask throughout an observation \citep{Rigb22}.

%\textit{JWST}'s pointing stability of $1$~mas is $\sim$6$-$7~times smaller than predictions, meaning smaller drifts in alignment behind the coronagraphic masks throughout an observation

The different contrasts as achieved by the ADI, RDI, and ADI+RDI subtractions allow for more concrete recommendations in observing structure for future programs. Most significantly, the improvement between RDI and ADI+RDI subtractions is negligible. Future observers will may be able to achieve their science goals with less telescope time by focusing on purely ADI or RDI subtraction strategies, however, a broader range of observations will be required to rule out ADI+RDI strategies entirely. Within 1$-$2\arcsec{} the ADI subtraction for MIRI F1550C, and to some extent F1140C, struggles to fully subtract the residual stellar PSF and an RDI based subtraction strategy should be preferred. Although RDI subtractions can provide improvements of up 5$-$10 times deeper contrast below 1\arcsec{} for some filters (see \fref{app:ccurves}), this technique requires a larger amount of observing time due to the cost-intensive nature of dithered reference observations. These time costs can be reduced by performing a smaller number of reference dithers, however, this will in turn reduce the achieved contrast. For example, by comparing subtractions using individual reference dithers from a nine-point dither strategy, \citet{Gira22} demonstrate a range of 2$-$10 times worse contrast at 1\arcsec{} compared to a subtraction combining all dithers in a PCA framework similar to that adopted for this work. A precise assessment of the trade space between contrast, observing time, dither pattern, and subtraction strategy for \textit{JWST} coronagraphy is beyond the scope of this work. Nevertheless, future observers should use simulation tools such as \texttt{PanCAKE} \citep{Cart21b} to identify the contrast performance necessary to meet their science goals, and adjust their PSF subtraction strategy accordingly to improve observational efficiency.

\rev{Similar NIRCam and MIRI observations have been performed as part of observatory commissioning and are detailed in \citet{Gira22}, \citet{Kamm22}, and \citet{Bocc22}. The observed contrast between this work and these commissioning efforts are in broad agreement, with the regime within $\sim$2\arcsec{} being contrast limited, and the regime beyond $\sim$2\arcsec{} being background limited. For the NIRCam F356W observation of HIP~65426 (0.5~Jy) in the background limited regime we reach sensitivities of $\sim$0.4~$\mu$Jy ($\sim$20 minutes exposure time), compared to $\sim$1~$\mu$Jy in \citet{Kamm22} for HD~114174 (G3IV, 2MASS~$K_{s}$=5.202) in the F335M filter (2.6~Jy, $\sim$55 minute exposure). For MIRI F1140C/F1150C observations of HIP~65426 (0.06/0.03~Jy) in the background limited regime we reach sensitivities of $\sim$2.7/3~$\mu$Jy ($\sim$30/120 minutes exposure time), compared to $\sim$9/30~$\mu$Jy in \citet{Bocc22} for HD~158165(K5V, 2MASS~$K_{s}$=4.704)/HD~163113(K0V, 2MASS~$K_{s}$=2.749) in the F1140C/F1150C filters (0.45/1.45~Jy, $\sim$75/150 minutes exposure time). Differences in these measured sensitivities are likely driven by the complex interplay of source spectral type, magnitude, and exposure time, and will be better understood in the context of a broader range of future JWST coronagraphic observations.}

Beyond this work, there is significant potential for the contrast performance for both NIRCam and MIRI to improve. \citet{Kamm22} have already demonstrated that the NIRCam bar masks can provide deeper contrasts at shorter separations, if a full 360\degree{}~field-of-view is not required. Similarly, in later cycles it may be possible to position a star at the ``NARROW'' bar mask position in an attempt to reduce the effective inner working angle (IWA) even further. The efficacy of RDI subtractions may improve with the use of a larger PSF library populated by on-sky observations across multiple programs. It may also be possible to perform an effective PSF subtraction using a much larger PSF library composed either entirely or partially of model PSFs. A particular opportunity for improvement will also come from \textit{JWST} program GO-02627 \citep{Ygou21}, which aims to estimate the on-sky instrumental aberrations that drive variations in the observed PSF structure with a model based phase retrieval algorithm.

\subsection{Mass Sensitivity}
Using the obtained contrast curves we also determine the detectable mass limits for our observations following the approach of \citet{Cart21a}. Briefly, we convert our contrast curves to magnitude sensitivity curves using the stellar magnitudes as described in Section \ref{sec:contrast_cal}, and then convert these to a mass sensitivity using an interpolation of the evolutionary models of \citet{Lind19} (BEX) and \citet{Phil20} (ATMO) assuming an age of 14$\pm$4~Myr \citep{Chau17} and distance of 107.49$\pm$0.40~pc \citep{Gaia20}. As in \citet{Cart21a} we select the chemical equilibrium, non-cloudy models to maintain model consistency across mass ranges. Clouds and disequilibrium chemistry likely play a significant role in sculpting the emission of sub-stellar atmospheres, however, an investigation into these effects is beyond the scope of this work.

Following the calculation of mass sensitivity curves we use the Exoplanet Detection Map Calculator \citep[\href{https://ascl.net/2010.008}{\texttt{Exo-DMC}},][]{MESS, Bonavita2013, ExoDMC}\footnote{\url{https://github.com/mbonav/Exo_DMC}} to estimate detection probability maps. In this case, we produce a population of synthetic companions with masses and semi-major axes from $0.1\,\rm{M\textsubscript{Jup}}$  to $100\,\rm{M\textsubscript{Jup}}$ and $0.1\,\rm{au}$ to 10,000~au respectively. The inclination is uniformly distributed in $\sin{(i)}$, the eccentricity is distributed using a Gaussian with $\mu=0$ and $\sigma=0.3$ (excluding negative eccentricities; \citealt{Hogg2010}), and all other orbital parameters are uniformly distributed. Implicit in this is an assumption that these synthetic companions do not necessarily have a similar inclination to \target{}. The resulting map takes into account the effects of projection when estimating the detection probability, and the probability for a potential companion to truly lie in the instrumental field of view.

%\texttt{Exo-DMC} is a python adaptation of the \texttt{MESS} \citep[Multi-purpose Exoplanet Simulation System,][]{MESS, Bonavita2013}, using a Monte Carlo approach to compare instrument detection limits with a simulated, synthetic population of planets around the target star to estimate the probability of detection of a companion of a given mass and semi-major axis. 

%For each point in the mass/semi-major axis grid, \texttt{Exo-DMC} generates a fixed number of sets of orbital parameters. 

% \texttt{Exo-DMC} allows for a high level of flexibility in terms of possible assumptions on the synthetic planet population to be used for the determination of the detection probability. 

%However, we choose to use \texttt{Exo-DMC}'s basic setup, with a flat logarithmic distribution for both mass and semi-major axis, in order to avoid introducing further biases. 

\begin{figure}[t]
    \centering
    \includegraphics[width=\columnwidth]{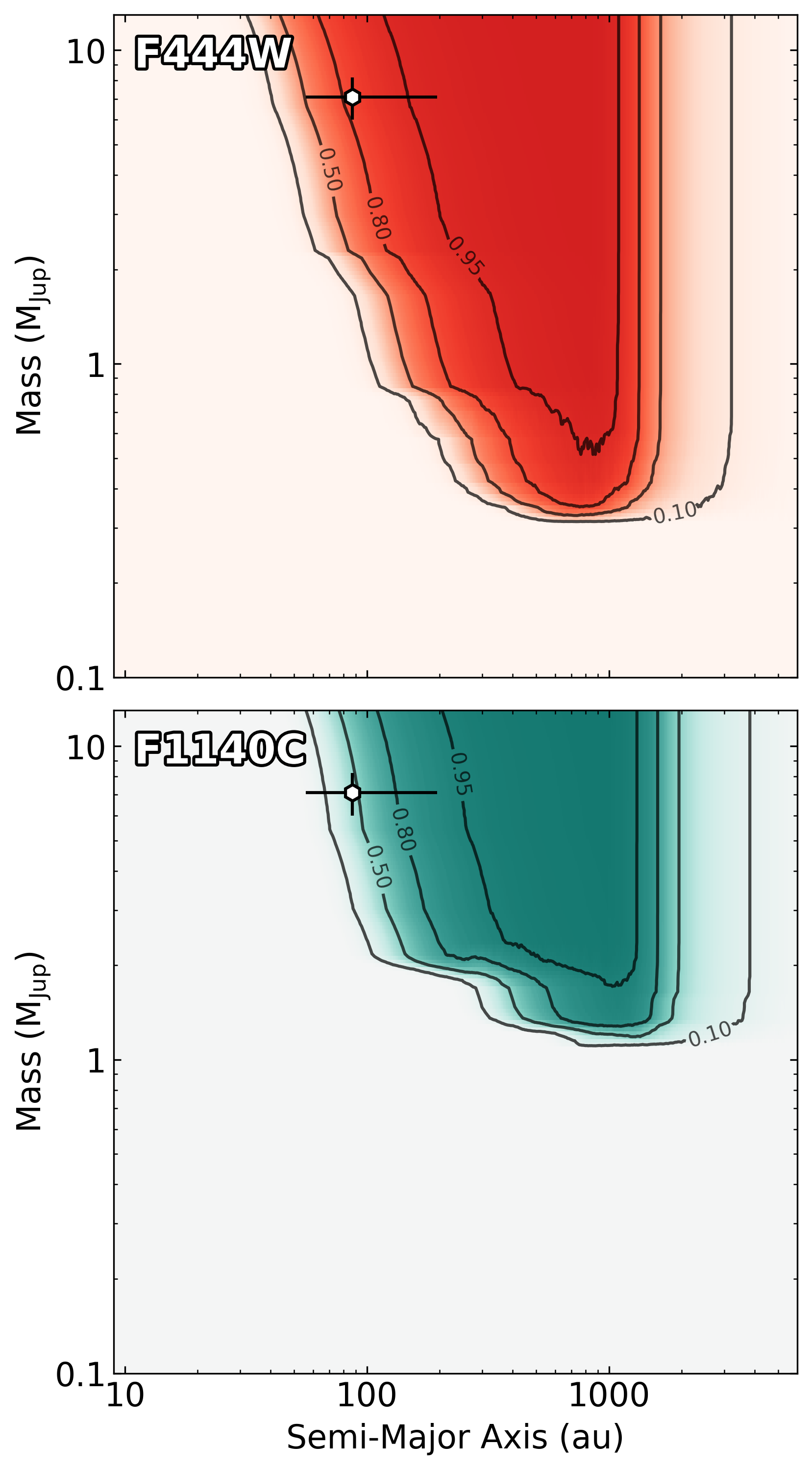}
    \caption{Detection probability maps as generated by \texttt{Exo-DMC} for the most sensitive NIRCam (F444W, top) and MIRI (F1140C, bottom) filters. Solid black contours signify the 10\%, 50\%, 80\%, and 95\% detection thresholds. \rev{The true location of \target{} is marked (hexagon) with errors taken from the astrometric fitting in Section \ref{sec:astrometry}, and evolutionary model analysis in Section \ref{sec:bolmass}.}}
    \label{fig:mass_sens}
\end{figure}

For these particular observations of HIP\,65426, we identify the F444W and F1140C filters as the most sensitive to the lowest mass companions for NIRCam and MIRI, respectively, and display their detection probability maps in \fref{fig:mass_sens}. Of the two, the F444W filter is the most sensitive, reaching a sub-Jupiter mass sensitivity from $\sim$150$-$2000~au at a 50\% probability, and a minimum sensitivity of $\sim$0.4~$M_\mathrm{Jup}$ ($\sim$350~K) $\sim$300$-$1500~au at a 50\% probability. In contrast, the F1140C is unable to reach sub-Jupiter mass sensitivity, and has a minimum sensitivity of 1.5~$M_\mathrm{Jup}$ ($\sim$550~K) from $\sim$150$-$2000~au. As we detect no sources within the observed field-of-view that have colours consistent with a planetary mass companion, we set equivalent limits on the presence of additional companions in the HIP\,65426 system. 

\rev{Despite the F444W observation being at a shorter wavelength, corresponding to a finer angular resolution, it does not probe significantly closer separations than the F1140C observation. This is driven by the competing influence of the larger IWA for the NIRCam MASK335R of $\sim$0.63\arcsec, compared to $\sim$0.36\arcsec{} for the MIRI FQPM1140C (both assumed at 50\% transmission radius). At the wider separations, the F444W observation is sensitive to the lower mass companions than the F1140C observation due to the lower thermal background, which is $\sim$100 times fainter at 4.5$\mu$m compared to 11.3$\mu$m \citep{Rigb22}\footnote{\href{https://jwst-docs.stsci.edu/jwst-general-support/jwst-background-model}{https://jwst-docs.stsci.edu/jwst-general-support/jwst-background-model}}.}

As discussed in \citet{Cart21a}, at a given distance A~stars are generally poor targets for detecting the lowest mass planets in terms of detection sensitivity, whereas M~stars are among the most favourable. Given the improved performance of \textit{JWST}, it is likely that for nearby targets within (or outside of) the M~star sample from \citet{Cart21a} it will be possible to detect Uranus and Neptune mass objects beyond $\sim$100$-$200~au, and Saturn mass objects beyond $\sim$10~au. Initial searches across a small sample of stars for sub-Jupiter mass objects will be performed as part of guaranteed time observations \citep{Schl17}. Furthermore, for the nearest targets we may be able to push these limits even further, with planned observations of $\alpha$~Cen~A aiming to be sensitive to $~$5~$R_\oplus$ companions from 0.5$-$2.5~au \citep{Beic20, Beic21}. 

%In \citet{Cart21a}, a sample of $\sim$60~M~stars provided a detection sensitivity of $<$0.2~$M_\mathrm{Jup}$ at 50\% probability. 

%This is an interesting contrast to the work shown in \citet{Cart21a}, of A stars in the nearby TW~Hya Association ($\sim$10~Myr, $\sim$60~pc \citealt{Kast97, Gagn18a}) and $\beta$~Pictoris Moving Group ($\sim$24~Myr, $\sim$30~pc \citealt{Zuck01, Gagn18a}). Whilst the F1140C sensitivity from this work is similar to that of \citet{Cart21a}, the F444W sensitivity has greatly improved from a minimum sensitivity of 1$-$2~$M_\mathrm{Jup}$ at a 50\% probability. As HIP\,65426 has a similar age of 14~Myr, this discrepancy is likely driven by the greater relative improvement of the F444W contrast over the F1140C contrast compared to pre-launch predictions (see \ref{sec:ach_contrast}, \fref{fig:all_contrasts}). 

\section{\target{} In Context}\label{sec:hip65426b_results}
The known companion \target{} is clearly detected in all seven of the observational filters using RDI, and all filters except the MIRI F1550C using ADI. These observations represent the first images of an exoplanet to be obtained with \textit{JWST}, and the first ever direct detection of an exoplanet beyond 5~$\mu$m.

\begin{figure*}
    \centering
    \includegraphics[width=\textwidth]{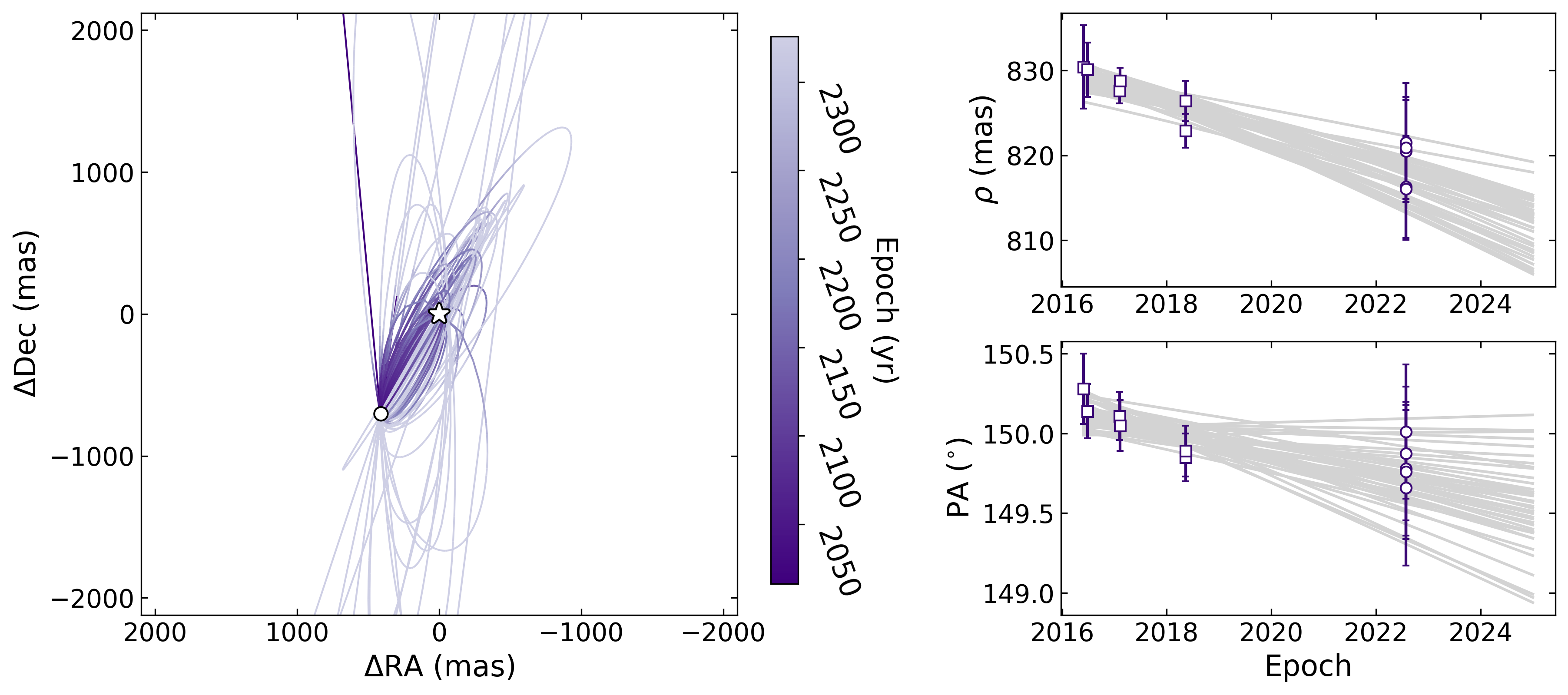}
     \includegraphics[width=\textwidth]{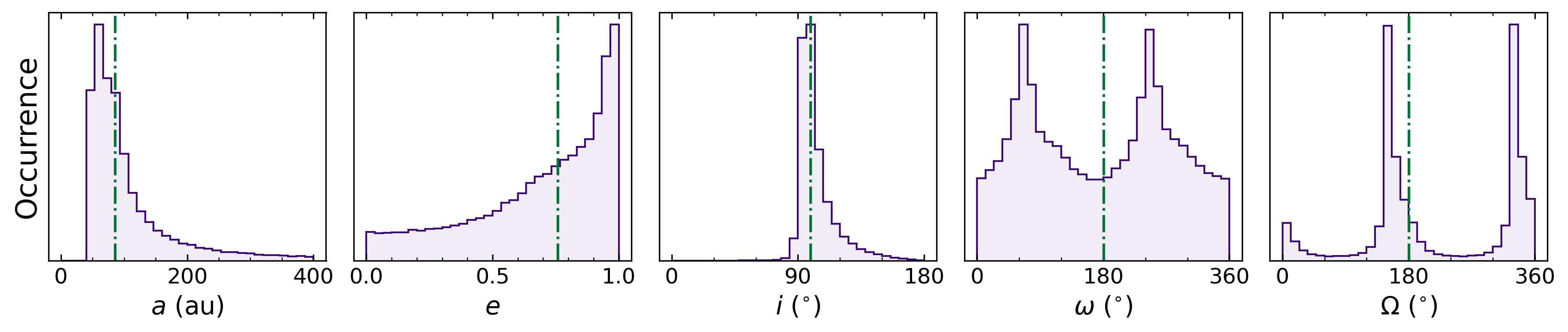}
    \caption{Orbital fitting of both \textit{JWST} NIRCam (this work) and SPHERE \citep{Chau17, Chee19} astrometric measurements of \target{} using \texttt{orbitize!} \citep{Blun17}. \textbf{\textit{Top Left:}} A random sample of 100 orbit models from the retrieved posterior (purple curves). The position of the planet (white circle) and the star (white star) are marked, and the epoch at a given position in an orbit is indicated by the colour bar. \textbf{\textit{Top Right:}} Separation and position angle versus epoch for both the SPHERE (squares) and \textit{JWST} NIRCam astrometry (circles). The same random sample of 100 orbits is also displayed (grey lines). \textbf{\textit{Bottom:}} Posterior distributions for the semi-major axis ($a$), eccentricity ($e$), inclination ($i$), argument of periastron ($\omega$), and position angle of nodes ($\Omega$). The 50\% quantile from these distributions (green dot-dashed line) are also indicated.  In this particular case, these additional \textit{JWST} observations do not significantly increase the constraints on the measured orbital parameters.}
    \label{fig:orbit}
\end{figure*}

\subsection{Astrometry}\label{sec:astrometry}
The measured astrometry in each of the observed filters is obtained from the ADI+RDI reduction and shown in \tref{tab:photastr}. Each of the measured uncertainties for the NIRCam astrometry are propagated with an additional uncertainty of 6.3~mas (0.1~pixels), and the uncertainties of the MIRI astrometry are propagated with an additional uncertainty of 10~mas (0.1~pixels) to account for the assumed precision of the absolute star centering as described in \ref{sec:alignment}. \rev{All NIRCam and MIRI astrometry are consistent within 1$\sigma$, and in combination the NIRCam and MIRI astrometry provide a measurement of the separation, $\rho$=819$\pm$6~mas, and the position angle, $\theta$=149.8$\pm$0.4\degree.} To compute these average values, we do not treat the NIRCam filters as independent and instead average both the quantities and their uncertainties. The absolute position of the star on the detector does not change significantly ($<$1 pixel) between NIRCam filters, and it is feasible that the measured position has a similar systematic offset in each filter (see Section \ref{sec:alignment}). As the dominant noise source for the NIRCam alignment is this systematic offset, it is not appropriate to propagate the uncertainties of the NIRCam astrometry in a typical fashion. 

We combine these new measurements with the existing astrometry from \citet{Chau17} and \citet{Chee19} to determine updated orbital parameters using the \texttt{orbitize!} package \citep{Blun20}. As in \citet{Chee19}, we exclude the NaCo epochs due to the inconsistency with the SPHERE epochs. Additionally, we exclude the MIRI epoch as the observations were taken just two weeks after the NIRCam observations and have larger uncertainties. \texttt{orbitize!} is initialised assuming one companion to the primary (\target{}), a total system mass of 1.97$\pm$0.046~$M_{\odot}$ \citep{Chau04}, and a parallax of 9.3031$\pm$0.0346~mas \citep{Gaia20}. Orbit generation is performed using the Orbits for the Impatient (OFTI) algorithm \citep{Blun20} until 100,000 possible orbits are identified. A random sample of 100 of the possible orbits along with posterior distributions for the entire sample are shown in \fref{fig:orbit}. \rev{We are able to constrain the semi-major axis, $a=86_{-\,31}^{+116}$~au, and the inclination, $i=99_{-\,6}^{+14}$ degrees (relative to the equatorial plane).} Additionally, as the motion of \target{} is primarily radial, solutions that place the line of nodes close to its position angle are preferred, and the position angle of nodes is also constrained to two possible solutions.

The addition of the NIRCam astrometry does not significantly improve the orbital constraints for \target{}, and all retrieved properties are consistent with those from \citet{Chee19} and \citet{Bowl20}. Although our eccentricity distribution more strongly favours higher eccentricities compared to \citet{Chee19}, it remains essentially unconstrained and should not be interpreted as evidence for a highly eccentric orbit for \target{}. However, if this high eccentricity is real, it would give credence to the scenario proposed in \citet{Marl19}, where \target{} initially formed through core accretion before being scattered to a wider separation by an additional companion. The ability of \textit{JWST} to provide high precision astrometry may improve with improvements to the measurement of the absolute star position in the NIRCam images, which in this case is limited by the precision with which we can locate the star center behind the coronagraphic mask (see Section \ref{sec:alignment}).

Separate to \textit{JWST}, \target{} has been observed using \textit{VLT} interferometry as part of the ExoGRAVITY program (\citealt{Laco20}, Program ID: 1104.C-0651), which has routinely demonstrated sub$-$mas astrometric precision (e.g., \citealt{Grav19, Laco21, Hink22b}) and has an even greater potential to improve upon our reported constraints.

\subsection{Photometry}\label{sec:photometry}
\begin{figure*}
    \centering
    \includegraphics[width=\textwidth]{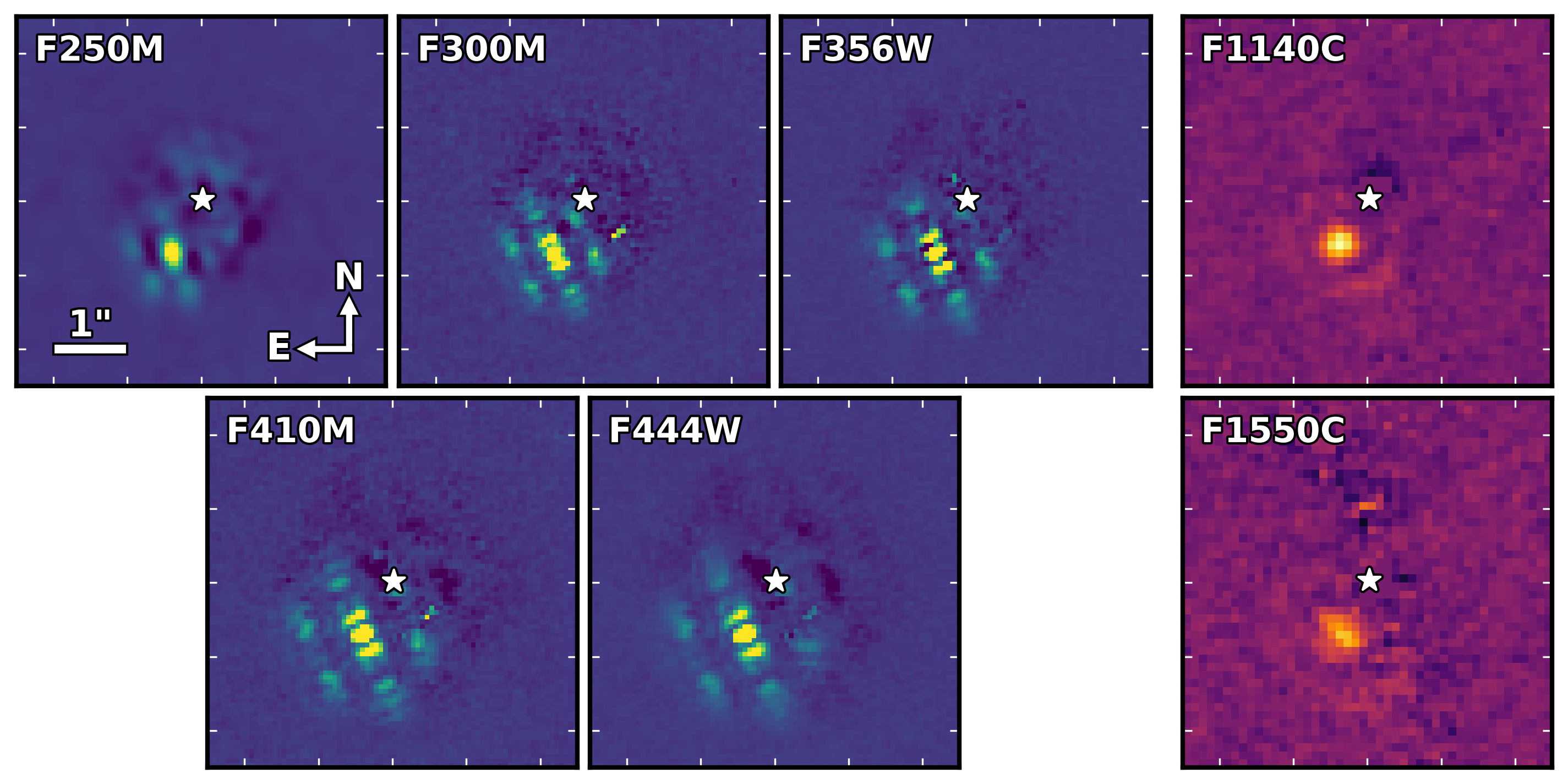}
    \caption{Images of the exoplanetary companion, \target{}, in all seven NIRCam and MIRI filters used in our observations. Each image is produced following an ADI+RDI KLIP subtraction of the residual stellar PSF. The measured position of the star is marked (white stars), and the orientation and pixel scales of all images are marked in the top left panel.} 
    \label{fig:adirdi_all}
\end{figure*}

\begin{figure*}
    \centering
    \includegraphics[width=\textwidth]{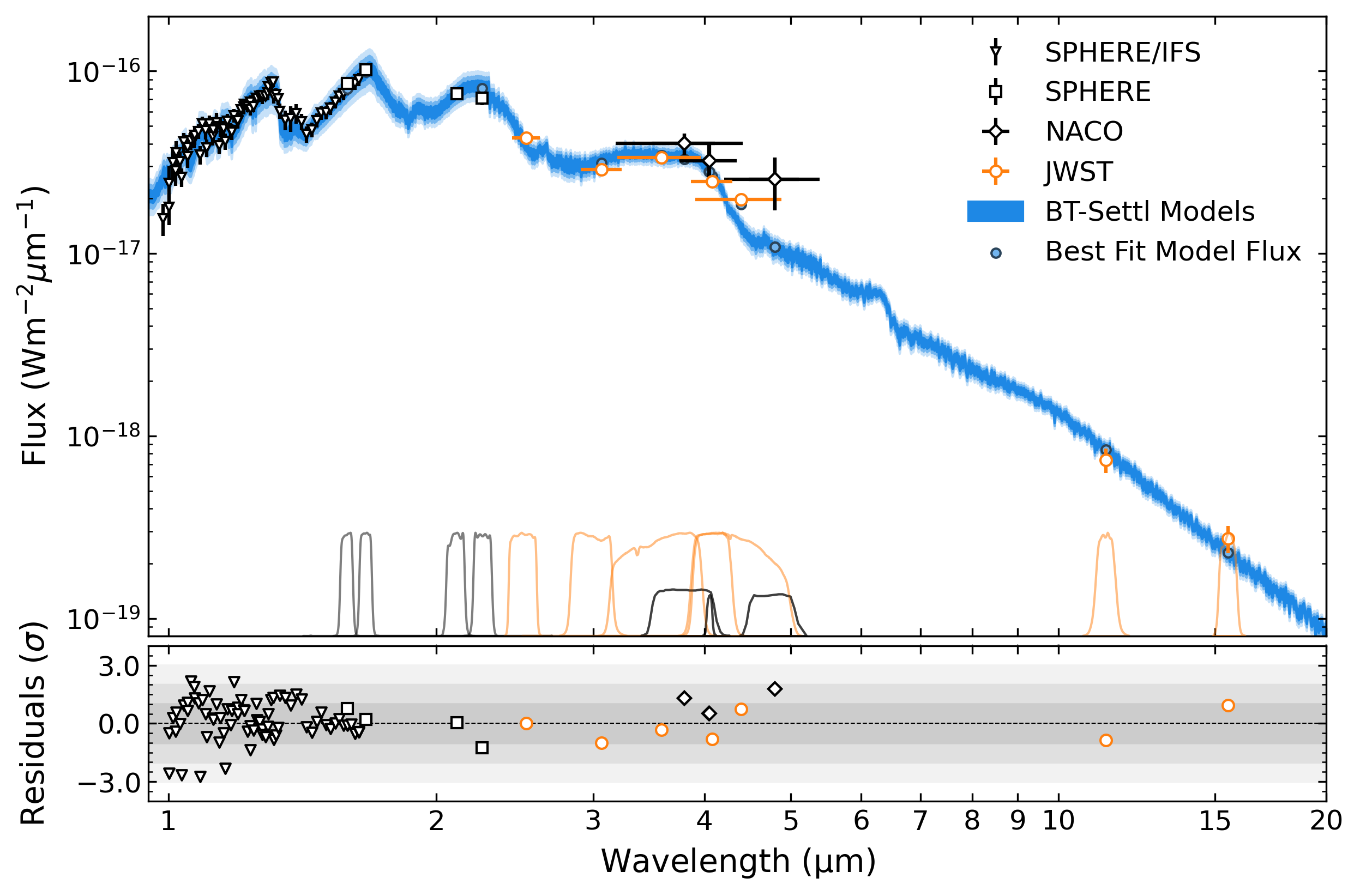}
    \caption{All existing spectroscopic and photometric observations of \target{} as obtained from SPHERE/IFS (triangles), SPHERE/IRDIS (squares), NaCo (diamonds), and \textit{JWST} (circles). \textbf{\textit{Top:}} Data are plotted alongside the 1, 2, and 3$\sigma$ confidence intervals obtained from fitting to a collection of \texttt{BT-SETTL} atmospheric forward models (blue shaded regions), and the model values in the photometric bandpasses (small blue circles). At 3$\sigma$, the best fit models occupy parameter ranges of $T_\mathrm{eff}$~=~1624$_{-15}^{+16}$~K, log($g$)~=~3.88$_{-0.08}^{+0.08}$~dex, and $R$~=~1.06$_{-0.05}^{+0.05}$~$R_\mathrm{Jup}$. Also plotted are the normalised filter throughput profiles for all photometric observations, with the NaCo throughputs scaled by a factor of 2 to improve clarity. \textbf{\textit{Bottom:}} Residuals of each data point relative to the best fit model in addition to 1, 2, and 3$\sigma$ regions (grey shading).}
    \label{fig:model_and_data}
\end{figure*}

As with the astrometry, the measured photometry in all of the observed filters is obtained from the ADI+RDI reduction and shown in \tref{tab:photastr}, the subtracted images of all of these filters is shown in \fref{fig:adirdi_all}, and the photometric data points themselves are shown in \fref{fig:model_and_data}. \rev{The measured contrast of the planet relative to the star ranges from 10.132~mag in the F250M filter to 8.029~mag in the F1550C filter}. Images for the ADI and RDI subtractions can be found in Appendix \ref{app:images}. Additional literature photometric measurements as shown in \fref{fig:model_and_data} are provided in Appendix \ref{app:phottable}. 

\begin{table*}
    \centering\setlength{\tabcolsep}{6pt}
    \begin{tabular}{cccccccc}
        \hline
        \hline
        \vspace{1mm}
        Filter & $\rho$ (mas) & $\theta$ (deg) & $m_\mathrm{*}$ (mag) & $\Delta$ (mag) & $\Delta_\mathrm{corr}$ (mag) & $m_\mathrm{b}$ (mag) & Flux (Wm$^{-2}\mu$m$^{-1}$) \vspace{1mm} \\
        \hline \vspace{-4mm}
        \\
        F250M & 822$\pm$7 & 149.7$\pm$0.5 & 6.783$\pm$0.054 & 10.132$\pm$0.032 & 10.132$\pm$0.057 & 16.915$\pm$0.083 & (4.29$\pm$0.33)$\times10^{-17}$ \\
        F300M & 821$\pm$6 & 149.8$\pm$0.4 & 6.766$\pm$0.046 & 9.829$\pm$0.027 & 9.829$\pm$0.056 & 16.595$\pm$0.076 & (2.89$\pm$0.20)$\times10^{-17}$ \\
        F356W & 816$\pm$6 & 150.0$\pm$0.4 & 6.767$\pm$0.048 & 8.980$\pm$0.015 & 8.980$\pm$0.055 & 15.747$\pm$0.074 & (3.36$\pm$0.23)$\times10^{-17}$ \\
        F410M & 816$\pm$6 & 149.8$\pm$0.4 & 6.765$\pm$0.051 & 8.734$\pm$0.019 & 8.734$\pm$0.055 & 15.499$\pm$0.077 & (2.49$\pm$0.18)$\times10^{-17}$ \\
        F444W & 820$\pm$6 & 149.9$\pm$0.4 & 6.764$\pm$0.054 & 8.703$\pm$0.015 & 8.703$\pm$0.055 & 15.467$\pm$0.078 & (1.97$\pm$0.13)$\times10^{-17}$ \\
        F1140C & 823$\pm$11 & 149$\pm$1 & 6.722$\pm$0.038 & 8.264$\pm$0.021 & 8.264$\pm$0.164 & 14.986$\pm$0.169 & (7.40$\pm$1.16)$\times10^{-19}$ \\
        F1550C & 836$\pm$15 & 149$\pm$1 & 6.766$\pm$0.072 & 8.029$\pm$0.039 & 8.029$\pm$0.167 & 14.705$\pm$0.182 & (2.74$\pm$0.46)$\times10^{-19}$ \\
        \hline
    \end{tabular}
    \caption{\textit{JWST} astrometry and photometry of \target{}, $m_\mathrm{*}$ corresponds to the stellar magnitude in each filter, and $\Delta_\mathrm{corr}$ corresponds to the relative magnitude following the propagation in uncertainties of a 5\% or 15\% absolute flux calibration accuracy for NIRCam and MIRI, respectively. The position angle ($\theta$) is provided from North through East, and all apparent magnitudes are relative to Vega.}
    \label{tab:photastr}
\end{table*}

\subsection{Bolometric Luminosity}
With the addition of \textit{JWST} NIRCam and MIRI photometric observations, the SED of \target{} is measured across $1~\mu\mathrm{m}$ to $15~\mu\mathrm{m}$. The measurements span the majority of its luminous wavelength range and enable a tight constraint on the bolometric luminosity of the planet.
 
To calculate the luminosity, a full SED was created by distributing the flux-density from photometric measurements over the effective bandwidth for each filter and using a model atmosphere to extrapolate beyond and interpolate between measured bands. Luminosity is then determined by integrating this semi-empirical SED over wavelength. 

Since all of the flux measured in the NIRCam/F410M photometry is accounted for in the F444W measurement, we used only the wider band for our analysis. We also added measurements from the literature at shorter wavelengths, including the SPHERE-IFS YH-band spectrum \citep{Chee19}, and SPHERE-IRDIS H3, K1, and K2-band photometry (\citealt{Chau17, Chee19}; also see Appendix \ref{app:phottable}).

To explore the dependence on the details of atmospheric model assumptions, we calculated the bolometric luminosity multiple times, using a broad range of models for interpolation and extrapolation of the SED. Atmospheric models spanned $T_{\mathrm{eff}}$ from 1200~K to 1900~K and $\log(g)$ spanning 3.5 to 5.5.  These models were drawn from three different grids, including two with different cloud implementations -- BT-SETTL \citep{Bara15}, and DRIFT-Phoenix \citep{Witt09} -- and the cloud-free Sonora-bobcat models \citep{Marl2021}. No matter which model we used to fill in the gaps between the measured portions of the SED, \lbol{} is always between $-$4.14 and $-$4.31. Consequently, the luminosity is constrained at the 0.17 dex level and the result is robust across all considered model atmospheres. In totality, the measured luminosity fraction ranges from 61$-$89\% dependant on the adopted model atmosphere.

\subsection{Estimates of Mass and other Companion Properties from Hot-Start Evolutionary Models}\label{sec:bolmass}
The mass of \target{} is estimated using a method similar to that described in \citet{Dupu2017}. We first built an interpolated grid of model luminosities as a function of age and mass, with 10,000 equally-spaced age values spanning from 5 to 30~Myr and 10,000 equally-spaced mass values spanning from 0.3 to 21~M$_\mathrm{Jup}$ using \texttt{scipy.interpolate.griddata} \rev{with cubic interpolation} in \texttt{python}. We adopted an age for \target{} of 14$\pm$4~Myr based on the Lower Centaurus-Crux age given in \citet{Chau17} and a measured bolometric luminosity uniformly distributed between  \lbol~=~$-$4.14~and~$-$4.31 from the previous section.

We then generated 1$\times$10$^6$ samples of age and mass from a Gaussian distribution in age around 14~Myr, with $\sigma$=4~Myr, and a uniform distribution in mass from 0.3 to 21~M$_\mathrm{Jup}$. For each age, mass sample, we then look up the corresponding model luminosity from the interpolated grid of model luminosities. For each age, mass sample, we accept the sample if the corresponding model luminosity is within the measured range of uniformly-distributed bolometric luminosities and reject the sample if it lies outside this range.
%We then calculate $\chi^{2}$ for each age, mass sample as:
%\begin{equation}
%    \chi^{2} = \frac{(L_{bol,model} - %L_{bol,measured})^2}{\sigma_{Lbol,measured}^2}
%\end{equation}
%As our uncertainty on the measured bolometric luminosity is asymmetric, we adopt the lower error bar for model luminosities less than the measured bolometric luminosity and the upper error bar for model luminosities above the measured bolometric luminosity.  We then convert our $\chi^{2}$ values to probabilities by normalizing by the minimum $\chi^{2}$ value among our 1$\times$10$^6$ samples:
%\begin{equation}
%    p = e^{-(\chi^2 - min(\chi^2))/2}.
%\end{equation}
%To select which samples to retain, we also draw a set of 1$\times$10$^6$ uniformly distributed numbers from 0 to 1.  We compare each sample with its corresponding uniformly distributed variate and retain the samples where the sample probability is greater than the uniformly distributed variate drawn for that sample. 
%While \citet{Dupu2017} found that multiple iterations were necessary before the procedure settled upon a stable range of accepted masses, we found that the range of accepted masses did not change with additional iterations.  
%We then rerun the procedure for the range of masses among the accepted samples, iterating until it settles upon a stable range of masses.

\begin{figure*}
    \centering
    \includegraphics[width=7.2in]{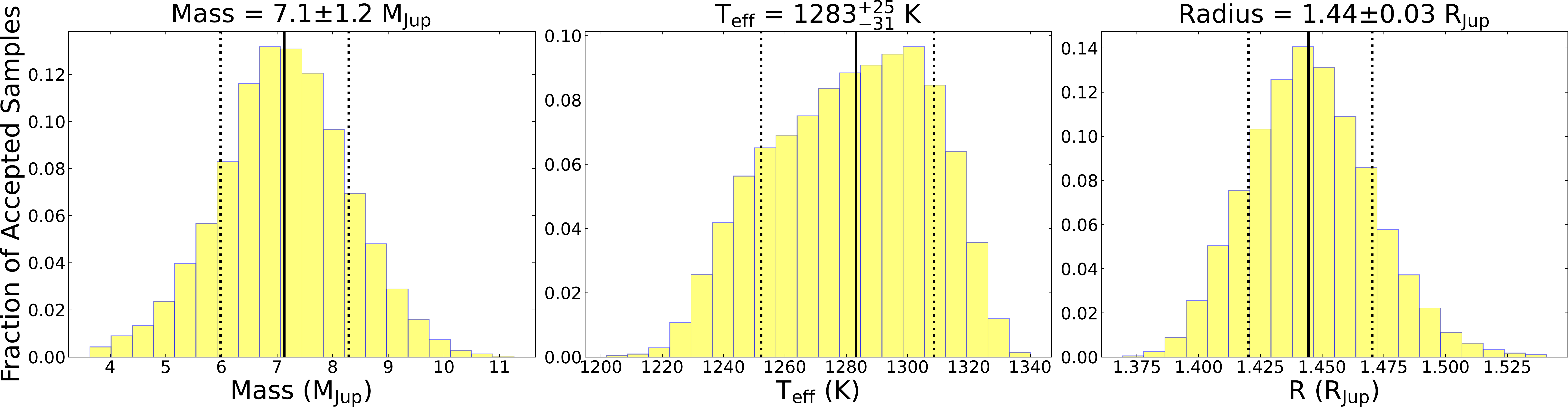}
    \caption{Histograms of the final sets of accepted model properties for the hybrid cloud grid from \citet{Saum2008}. The median value for each property is shown as a solid black line, with the 68\% confidence region falling between the two dotted black lines.}
    \label{fig:massranges}
\end{figure*}

We implemented this procedure using the hybrid cloud grid from \citet{Saum2008}. Given that this is a dusty, young red object, we expect the \citet{Saum2008} models, which take clouds into account, to provide the most reliable estimates of the properties of these objects among the model choices available.  To sample the corresponding effective temperatures, surface gravities, and radii corresponding to our accepted mass values, we built interpolated grids of model effective temperatures, surface gravities, and radii with the same spacing in age and mass as for the interpolated grid of luminosities, then looked up the corresponding property in the appropriate table for each accepted age, mass sample.  Histograms of the final set of accepted masses, effective temperatures, surface gravities, and radii for each model are shown in Fig.~\ref{fig:massranges}.  The best value of each parameter was taken as the median of the accepted distribution, with error bars given by the 68\% confidence interval as calculated from the histogram of each distribution.  \rev{We find a mass of 7.1$\pm$1.2 M$_\mathrm{Jup}$, radius of 1.44$\pm$0.03 R$_\mathrm{Jup}$ (which therefore imply a surface gravity log(g) = 3.93$^{+0.07}_{-0.09}$), and effective temperature T$_\mathrm{eff}$ = 1283$^{+25}_{-31}$ K.}

\vfill\null
\subsection{Atmospheric Forward Model Comparison}\label{sec:modelfitting}
To explore the atmospheric properties of \target{} we performed a forward modelling analysis using the tool \texttt{ForMoSA} \citep{Petr20}, which compares spectroscopic and/or photometric data with grids of precomputed synthetic spectra. The code is based on the nested sampling algorithm \citep{Skil04}, a Bayesian inversion method, that allows a global exploration of the parameter space provided by the grid. In this work, we limited our analysis to the \texttt{BT-SETTL} grid (\texttt{CIFIST} version, \citealt{Alla12, Bara15}) that accounts for convection using mixing-length theory, and works at hydrostatic, radiative-convective, and chemical equilibrium. 

%The code models the condensation, coalescence, and mixing of 55 types of grains, with abundances as determined by comparing the timescales of these different processes at each layer of the atmosphere, and then calculates the emitted flux by processing the radiative transfer with the \texttt{PHOENIX} code \citep{Haus97, Alla01}. This grid allows an exploration of the $T_\mathrm{eff}$ (from 1200 K to 7000 K) and the log($g$) (from 2.5 to 5.5 dex) with a solar metallicity. The input solar abundances are taken from \cite{Caff11}. 

For the fit, we used a data set composed of the low-resolution spectra (R$_{\lambda}\sim$54) between 1.00 and 1.65 $\mu$m provided by: \textit{VLT}/SPHERE-IFS \citep{Chau17}, \textit{VLT}/SPHERE-IRDIS $H_2$ (1.58 $\mu$m), $H_3$ (1.66 $\mu$m), $K_1$ (2.11 $\mu$m), and $K_2$ (2.25 $\mu$m) photometry \citep{Chee19}, \rev{NaCo $L'$ (3.77 $\mu$m), $N\!B4.05$ (4.06$\mu$m), and $M'$ (4.76 $\mu$m) photometry}, and our new \textit{JWST} NIRCam and MIRI photometry. We first adapted the \texttt{BT-SETTL} synthetic spectra to our data, by reducing their spectral resolution to that of SPHERE-IFS and calculating the synthetic photometric flux at each bandpass using throughputs as obtained from \sklip{} for the \textit{JWST} data, and from the SVO filter service for all other data\footnote{\href{http://svo2.cab.inta-csic.es/theory/fps/}{http://svo2.cab.inta-csic.es/theory/fps/}} \citep{Rodr12,Rodr20}. We then defined flat priors on the $T_\mathrm{eff}$ and the log($g$) according to the limits of the grid, and applied nested sampling to estimate the posterior distributions of these two parameters. We also add the radius, $R$, to the list of the parameters explored by the nested sampling. At each iteration, a radius is picked randomly (uniform prior), and a dilution factor $C_\mathrm{K}$ = ($R$/$d$)$^{2}$ is calculated and multiplied to the model, where $d$ is the distance to the object (107.49~pc).

\begin{figure*}
    \centering
    \includegraphics[width=0.75\textwidth]{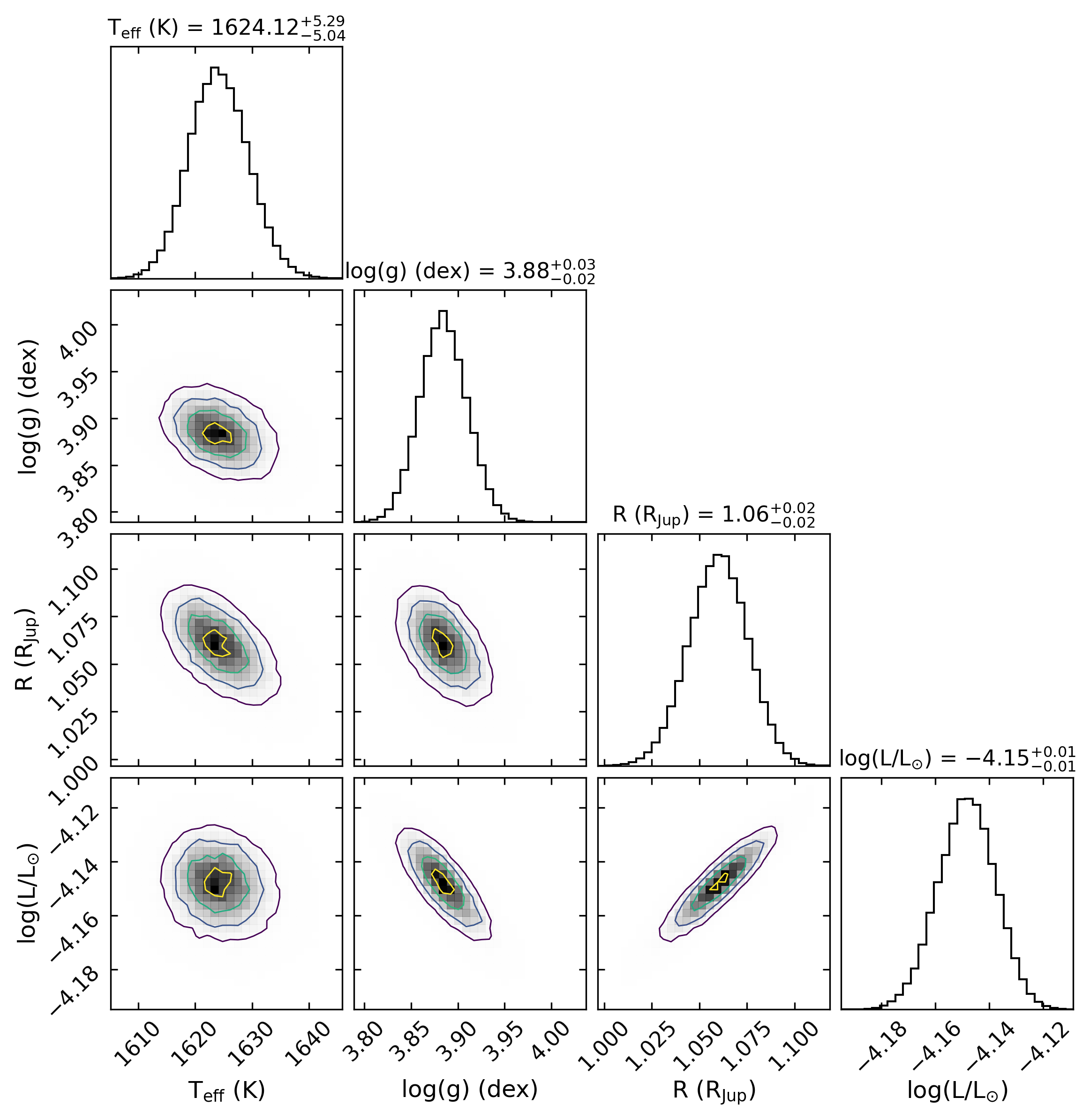}
    \caption{Posterior distributions for the BT$-$Settl atmospheric model fitting to both \textit{JWST} and $VLT$/SPHERE observations of \target{}. Best fit values and 1$\sigma$ uncertainties are indicated, however, these should be interpreted as the model phase space that fits these data, and not the precision to which these properties can be empirically measured.}
    \label{fig:model_post}
\end{figure*}

The best fit models to our data combined with the existing SPHERE and NaCo data are displayed in \fref{fig:model_and_data}, alongside posteriors in \fref{fig:model_post}. \rev{We estimate $T_\mathrm{eff}$~=~1624$_{-15}^{+16}$~K, log($g$)~=~3.88$_{-0.08}^{+0.08}$~dex, and $R$~=~1.06$_{-0.05}^{+0.05}$~$R_\mathrm{Jup}$. From the $T_\mathrm{eff}$ and the radius, we apply the Stefan-Boltzmann law and estimate a bolometric luminosity of \lbol=$-$4.15$^{+0.03}_{-0.03}$, and from the log($g$) and $R$ we estimate a mass of $M$=3.29$\pm$0.33~$M_\mathrm{Jup}$.} By comparing the integrated flux of the best fit model across all wavelengths, and the integrated flux between the shortest and longest wavelength observations, we determine that these observations span $\sim$97\% of \target's luminous range. These results are also in agreement with a similar \texttt{BT-SETTL} model fitting procedure to \textit{VLT}/SINFONI data of \target{} (see \citealt{Petr21}). The uncertainties of all parameters are given at 3$\sigma$, however, we emphasise that they do not necessarily describe our confidence in the true planetary properties and are better considered as the model phase space that best fits our data. 

The precision on these measurements is primarily driven by the SPHERE/IFS data, however, we do note some differences that result from the addition of the \textit{JWST} data. Specifically, when fitting just the SPHERE and NaCo data in isolation, we obtain \rev{$T_\mathrm{eff}$~=~1619$_{-17}^{+18}$~K, log($g$)~=~3.85$_{-0.09}^{+0.10}$~dex, $R$~=~1.10$_{-0.09}^{+0.09}$~$R_\mathrm{Jup}$ and \lbol~=~$-$4.12$^{+0.06}_{-0.06}$}, again with uncertainties given at 3$\sigma$. Therefore, the \textit{JWST} data improve the precision of the radius and bolometric luminosity by a factor of $\sim$2, but do not significantly improve the precision on the temperature and surface gravity. 

%We also note offsets between the best fit values between these two fits of up to 3$\sigma$ (using the uncertainties from the combined fit), which suggests that the selected models do not accurately capture these two regions of wavelength space simultaneously. 

%The uncertainty of all parameters are given at 3$\sigma$, however, we emphasise that they do not necessarily describe our confidence in the true planetary properties and are better considered as the model phase space that best fits our data. 

The atmospheric forward model fit yields a luminosity within the bolometric luminosity range of $-$4.14 to $-$4.31 that was found from combining SED measurements with models in the regions not covered by the SED. However, the effective temperatures and radii found using the atmospheric forward model are considerably in tension with predictions from evolutionary model fits to the measured bolometric luminosity range (see Section \ref{sec:bolmass}). In particular, to obtain similar bolometric luminosity values, the forward models favour higher effective temperatures ($\sim$1600~K) and smaller radii ($\sim$1.06~$R_\mathrm{Jup}$) compared to the evolutionary models ($\sim$1300~K, $\sim$1.44~$R_\mathrm{Jup}$). In fact, the atmospheric forward model used here corresponds to an unphysically small radius for an exoplanet which may still be contracting at these young ages. Thus, we consider the effective temperature and radius predictions from the evolutionary models to be more robust here. 

The tension we find between the atmospheric models and evolutionary models is well-documented in the community; atmospheric models have a long history of requiring unphysically small radii and high effective temperatures to fit spectroscopy \citep[see~e.g.,~][]{Maro08, Pati12}.  This stems from the different approaches and fundamental parameters underpinning evolutionary and atmospheric modelling techniques.

%The BT-Settl atmospheric models considered here are radiative-convective equilibrium, non-gray plane-parallel models that consider the radiative and convective energy transport through the atmosphere \citep{Alla12}.  
\rev{Atmospheric models produce model spectra as a function of effective temperature, $T_\mathrm{eff}$, surface gravity, $g$, and composition, irrespective of the mass or age of the object modelled.  When fitting atmospheric model spectra to observed spectra, the value of the radius necessary to produce the observed luminosity of the object can be derived if the distance to the object is known. From the radius and the best-fit model surface gravity, a mass value can be derived as well. However, these are not fundamental parameters of the model, but rather extrapolated quantities.} 
%For realistic modeling of brown dwarfs and giant exoplanets, such models must also include the effects of clouds and condensates on the atmosphere \citep{Saum2008}.

\rev{In contrast, evolutionary models couple a similar atmospheric model with a stellar-like interior model, solving for hydrostatic equilibrium, mass conservation, and assuming conservation of energy in shells within the planet \citep{Saum2008}.  
%We expect young giant planets to have fully convective interiors; evolutionary models match the entropy of the interior to the entropy of the convective bottom of the atmospheric model used to describe the eventual radiative transport of energy out of the atmosphere.  
The interior model has fundamental parameters of mass and age; running the model until radiative-convective equilibrium is reached yields a bolometric luminosity and radius.  
%The adjoining atmospheric model then yields an effective temperature and surface gravity. 
Mass and radius here are then properties of the model, as opposed to the case of the atmospheric models where they are derived quantities dependent in particular on the fit for $T_\mathrm{eff}$.} 

% In an observational context, we then fit atmospheric model spectra to observed infrared spectra / SEDs to yield estimates of T$_\mathrm{eff}$, $g$, and flux.  The radius of the object can then be estimated if the distance to the planet is known.  By themselves, atmospheric models do not yield information on the fundamental properties of a planet, namely its mass and age.

%   The SM hybrid models we use here couple a cloudy atmospheric model for spectral types expected to have dusty, silicate clouds (e.g. L dwarfs) and a clear atmospheric model for spectral types expected to have clear atmospheres (e.g. T dwarfs).  Thus, for every age, mass pair, there is a corresponding atmospheric model spectrum which has been integrated to yield the bolometric luminosity at that age and mass.

\rev{When fitting evolutionary models to observations, we fit directly to the bolometric luminosity. In this work the bolometric luminosity is measured to very high accuracy, as we are integrating over many wavelength bins. Thus, we find a robust fit to other fundamental properties such as mass, surface gravity, and $T_\mathrm{eff}$ from evolutionary models. However, if we considered the ``best-fit'' model spectrum corresponding to the parameters ($T_\mathrm{eff}$, log($g$)) of the best evolutionary model, it would poorly fit the observed spectrum. In contrast, the atmospheric model fits to the spectra involves directly fitting over many wavelength bins. Uncertainties in cloud parameterisation in current models pushes these fits to higher temperatures to explain the observed spectral features; as a result, to balance out to the measured luminosity of the object at its artificially high $T_\mathrm{eff}$, the radius derived from the spectral fit is pushed to implausibly low values. As a consequence of the small radius, the mass estimate from the atmospheric models is also unphysically low.} 

\rev{Further \textit{JWST} observations across a broad diversity of exoplanets and/or brown dwarfs will be able to empirically constrain these model discrepancies as a function of properties such as temperature, mass, and age, and may in turn uncover their precise origin and extent. Until then, the competing benefits and drawbacks between atmospheric and evolutionary mean that they are best considered in tandem as opposed to in isolation.} 

\subsection{Future Work}
There is a range of additional investigation that can be performed on the data presented here that is worth highlighting, but ultimately falls outside the scope of this work.  

Most importantly, it is possible to perform the atmospheric forward model fitting procedure shown here in Section \ref{sec:modelfitting} with a wide range of state-of-the-art models (e.g., ATMO, \citet{Trem15, Phil20}; Exo-REM, \citet{Char18}; Sonora, \citet{Kara21}), each with their own treatment for the effects of clouds and atmospheric chemistry. Additionally, atmospheric fitting can also be performed using retrieval techniques \citep[e.g.,][]{Moll20, Gonz21}. Divergences in the measured planetary properties between these models are expected, and a more complete analysis in the context of the relative assumptions of each model will greatly improve our understanding of the true properties of \target{}. 

The precision of the 3$-$5$\mu$m data may be sufficient to provide constraints on the relative atmospheric abundances of CH$_4$ and CO, which can be impacted by disequilibrium chemistry \citep{Zahn14,Mile20}. \rev{The F1140C photometry falls slightly under the best fit model, albeit only at $\sim$1$\sigma$, and may be indicative of absorption by small silicate dust grains \citep{Cush06, Suar22, Mile23}. Similarly, the F1550C photometry falls $\sim$1$\sigma$ above the best fit model, and may be sensitive to circumplanetary disk emission \citep{Ster04, Stol20b}.}

% The F1140C photometry falls slightly under the best fit model, albeit only at $\sim$1$\sigma$, and may be indicative of absorption by small silicate dust grains \citep{Cush06, Suar22, Mile23}. Finally, while the F1550C photometry is relatively insensitive to the choice of atmospheric model, it may be sensitive to circumplanetary disk emission \citep{Ster04, Stol20b}.

%%%%%%%%%%%%%%%%%%%%%%%%%%%%%%%%%%%%%%%%%%%%%%%%%%%%%%%%%%%%%%%%%%%%%%%%%%%%%%%%%%%%%%%%%%%%%%%%%%%%%%%%%%%%%%%%%%%%%%%%%%%%%%
% CONCLUSION %
%%%%%%%%%%%%%%%%%%%%%%%%%%%%%%%%%%%%%%%%%%%%%%%%%%%%%%%%%%%%%%%%%%%%%%%%%%%%%%%%%%%%%%%%%%%%%%%%%%%%%%%%%%%%%%%%%%%%%%%%%%%%%%
\section{Conclusion}\label{sec:conclusion}
In this work we present the first ever scientific observations using the \textit{JWST} high-contrast imaging modes of both NIRCam from 2$-$5~$\mu$m, and MIRI from 11$-$16~$\mu$m. The known exoplanet companion, \target{}, is clearly detected in all seven observational filters, representing the first ever direct detection of an exoplanet beyond 5~$\mu$m. These observations provide a variety of insights into: a) the performance and best practices of \textit{JWST} high-contrast imaging, and b) the properties of the \target{} system, which we summarise below: 

\begin{itemize}

\item \textit{\textbf{Contrast:}} \textit{JWST} is exceeding its anticipated contrast performance for both NIRCam and MIRI coronagraphy by up to a factor of 10 in the contrast limited regime (see Section \ref{sec:ach_contrast}). For the contrasts achieved, we are sensitive to sub-Jupiter companions with masses as small as 0.3$M_\mathrm{Jup}$ beyond separations of $\sim$100~au. Furthermore, for more optimal targets such as young, nearby M~stars it is highly likely that both NIRCam and MIRI will be sensitive to sub-Saturn mass objects beyond $\sim$10~au \citep{Cart21a}. 

\item \textit{\textbf{Subtraction Strategy:}} For these data at small separations $<$2\arcsec{}, the best contrast is obtained using a small-grid dither RDI subtraction strategy for both NIRCam and MIRI. Additionally, an ADI+RDI subtraction does not significantly improve the measured contrast compared to RDI. For the MIRI F1550C observations in particular, we were unable to recover \target{} using ADI alone. At wider separations however, the observational efficiency of ADI may make it preferable to RDI. These conclusions may aid future observers in selecting their PSF subtraction strategy, although we emphasise that a clearer understanding of whether they apply under all circumstances will require the analysis of a broader range of \textit{JWST} coronagraphic observations. 

\item \textit{\textbf{Photometry:}} These photometric observations of \target{} provide exquisite sensitivity at a precision of $\sim$7\% for NIRCam and $\sim$16\% for MIRI. Furthermore, prior to propagation of the uncertainty in the stellar flux ($\sim$5\%), and the current absolute flux calibration accuracy (5/15\% for NIRCam/MIRI), the uncertainty in the measured relative flux is even smaller at $\sim$2\% for both NIRCam and MIRI. These measurements are a significant step forward from ground-based observations from 3$-$5~$\mu$m, which have comparative uncertainties of $\sim$13$-$32\% for \target{}, and are restricted to particular wavelength regions due to telluric contamination. With this improved precision we will be able to constrain directly imaged exoplanet atmospheres in much greater detail, in addition to more complex effects such as variability, disequilibrium chemistry, and the emission of circumplanetary material. 

\item \textit{\textbf{Atmospheric Model Fitting:}}
Using a \texttt{BT-SETTL} atmospheric forward model we are able to fit all data, in addition to the majority of ground-based observations to within 2$\sigma$. \rev{This agreement provides precise constraints on the $T_\mathrm{eff}$~=~1624$_{-15}^{+16}$~K, log($g$)~=~3.88$_{-0.08}^{+0.08}$~dex, $R$~=~1.06$_{-0.05}^{+0.05}$~$R_\mathrm{Jup}$, and \lbol=$-$4.15$^{+0.03}_{-0.03}$. Compared to a fit excluding the \textit{JWST} data, this corresponds to a factor of $\sim$2 improvement in the precision of the radius and bolometric luminosity.} Despite the excellent model agreement, both the temperature and unphysically small radius are in disagreement with the values obtained from the evolutionary models, further emphasising a long standing tension for this class of objects. 

\item \textit{\textbf{Empirical Bolometric Luminosity:}} 
\rev{As \textit{JWST} offers a uniquely broad spectral coverage in comparison to ground-based instruments, we are able to obtain a very precise measurement of the bolometric luminosity of \target{} that is constrained between a \lbol=$-$4.14 to $-$4.31, irrespective of the model atmosphere adopted for the wavelengths not covered by observations. In combination with evolutionary models, this provides tight constraints on the properties of \target{} with $M$=7.2$\pm$1.1~$M_\mathrm{Jup}$, $T_\mathrm{eff}$=1283$^{+25}_{-31}$~K, and $R$=1.44$\pm$0.03~$R_\mathrm{Jup}$.} Given the achieved sensitivity, similar \textit{JWST} observations will facilitate this analysis for a broader range of PMCs than ever before and provide comparable constraints on their bolometric luminosities, and therefore mass. These measurements will in turn be valuable for investigating discrepancies between atmospheric and evolutionary models of exoplanets. 
\end{itemize}

In conclusion, the observations reported here from our ERS program (ERS-01386, \citealt{Hink22}) demonstrate that \textit{JWST} provides an transformative opportunity to study exoplanets through high-contrast imaging. Beyond this work, we also highlight existing and future publications from our program of: 3$-$16~$\mu$m NIRCam and MIRI coronagraphy of the circumstellar disk HD\,141569\,A (Millar-Blanchaer et al. in preparation, Choquet et al. in preparation); NIRSpec and MIRI spectroscopy from 1$-$28~$\mu$m of the PMC VHS\,J1256\,b \citep{Mile23} and NIRISS AMI observations of HIP\,65426 at 3.8~$\mu$m (Ray et al. in preparation, Sallum et al. in preparation).

%%%%%%%%%%%%%%%%%%%%%%%%%%%%%%%%%%%%%%%%%%%%%%%%%%%%%%%%%%%%%%%%%%%%%%%%%%%%%%%%%%%%%%%%%%%%%%%%%%%%%%%%%%%%%%%%%%%%%%%%%%%%%%
% ACKNOWLEDGEMENTS %
%%%%%%%%%%%%%%%%%%%%%%%%%%%%%%%%%%%%%%%%%%%%%%%%%%%%%%%%%%%%%%%%%%%%%%%%%%%%%%%%%%%%%%%%%%%%%%%%%%%%%%%%%%%%%%%%%%%%%%%%%%%%%%
\section{Acknowledgements}
We are truly grateful for the countless hours that thousands of people have devoted to the design, construction, and commissioning of \textit{JWST}. ALC acknowledges the significant harm caused to members of the LGBTQIA+ community in the Department of State and NASA, while under the leadership of James Webb as Under Secretary of State and NASA Administrator, respectively. This project was supported by a grant from STScI (\textit{JWST}-ERS-01386) under NASA contract NAS5-03127. This work benefited from the 2022 Exoplanet Summer Program in the Other Worlds Laboratory (OWL) at the University of California, Santa Cruz, a program funded by the Heising-Simons Foundation. ALC and this work have greatly benefited from ExoExplorers, which is sponsored by the Exoplanets Program Analysis Group (ExoPAG) and NASA’s Exoplanet Exploration Program Office (ExEP). This work has made use of the SPHERE Data Centre, jointly operated by OSUG/IPAG (Grenoble), PYTHEAS/LAM/CeSAM (Marseille), OCA/Lagrange (Nice), Observatoire de Paris/LESIA (Paris), and Observatoire de Lyon/CRAL, and is supported by a grant from Labex OSUG@2020 (Investissements d’avenir – ANR10 LABX56). SP acknowledges the support of ANID, -- Millennium Science Initiative Program -- NCN19\_171. MBo acknowledges support in France from the French National Research Agency (ANR) through project grant ANR-20-CE31-0012. This project has received funding from the European Research Council (ERC) under the European Union's Horizon 2020 research and innovation programme (COBREX; grant agreement n\degree{} 885593; EPIC, grant agreement n\degree{} 819155)). All the {\it JWST} presented in this paper were obtained from the Mikulski Archive for Space Telescopes (MAST) at the Space Telescope Science Institute and can be accessed via: \dataset[10.17909/2bdf-3p61]{http://dx.doi.org/10.17909/2bdf-3p61}. This research has also made use of the NASA Astrophysics Data System Bibliographic services and the \texttt{python} \citep{python} modules listed below.

\software{\texttt{NumPy} \citep{numpy}, \texttt{matplotlib} \citep{matplotlib}, \texttt{Astropy} \citep{astropy1, astropy2}, \texttt{SciPy} \citep{scipy}, \texttt{scikit-image} \citep{scikitimage}, \texttt{synphot} \citep{synphot}, \texttt{emcee} \citep{Fore13}, \texttt{MultiNest} \citep{Fero09, Buch14}, \texttt{PHOENIX} \citep{Alla12}, \texttt{webbpsf} \citep{Perr12, Perr14}, \texttt{PanCAKE} \citep{Cart21b}, \texttt{orbitize!} \citep{Blun17,Blun20}, \texttt{Exo-DMC} \citep{Bonavita2013, ExoDMC}, \texttt{pyKLIP} \citep{Wang15}, \sklip{} \citep{Kamm22}.}
%%%%%%%%%%%%%%%%%%%%%%%%%%%%%%%%%%%%%%%%%%%%%%%%%%%%%%%%%%%%%%%%%%%%%%%%%%%%%%%%%%%%%%%%%%%%%%%%%%%%%%%%%%%%%%%%%%%%%%%%%%%%%%
% REFERENCES %
%%%%%%%%%%%%%%%%%%%%%%%%%%%%%%%%%%%%%%%%%%%%%%%%%%%%%%%%%%%%%%%%%%%%%%%%%%%%%%%%%%%%%%%%%%%%%%%%%%%%%%%%%%%%%%%%%%%%%%%%%%%%%%
\bibliography{ers_hip65426}{}
\bibliographystyle{aasjournal}

%%%%%%%%%%%%%%%%%%%%%%%%%%%%%%%%%%%%%%%%%%%%%%%%%%%%%%%%%%%%%%%%%%%%%%%%%%%%%%%%%%%%%%%%%%%%%%%%%%%%%%%%%%%%%%%%%%%%%%%%%%%%%%
% APPENDIX %
%%%%%%%%%%%%%%%%%%%%%%%%%%%%%%%%%%%%%%%%%%%%%%%%%%%%%%%%%%%%%%%%%%%%%%%%%%%%%%%%%%%%%%%%%%%%%%%%%%%%%%%%%%%%%%%%%%%%%%%%%%%%%%
\renewcommand\cleardoublepage{%
 \clearpage
 \ifodd\value{page}\else\stepcounter{page}\fi
}
\appendix
\vspace{-8mm}
\section{Subtracted Images}\label{app:images}\vspace{-7mm}
\begin{figure*}[b]
    \centering
    \includegraphics[width=0.97\textwidth]{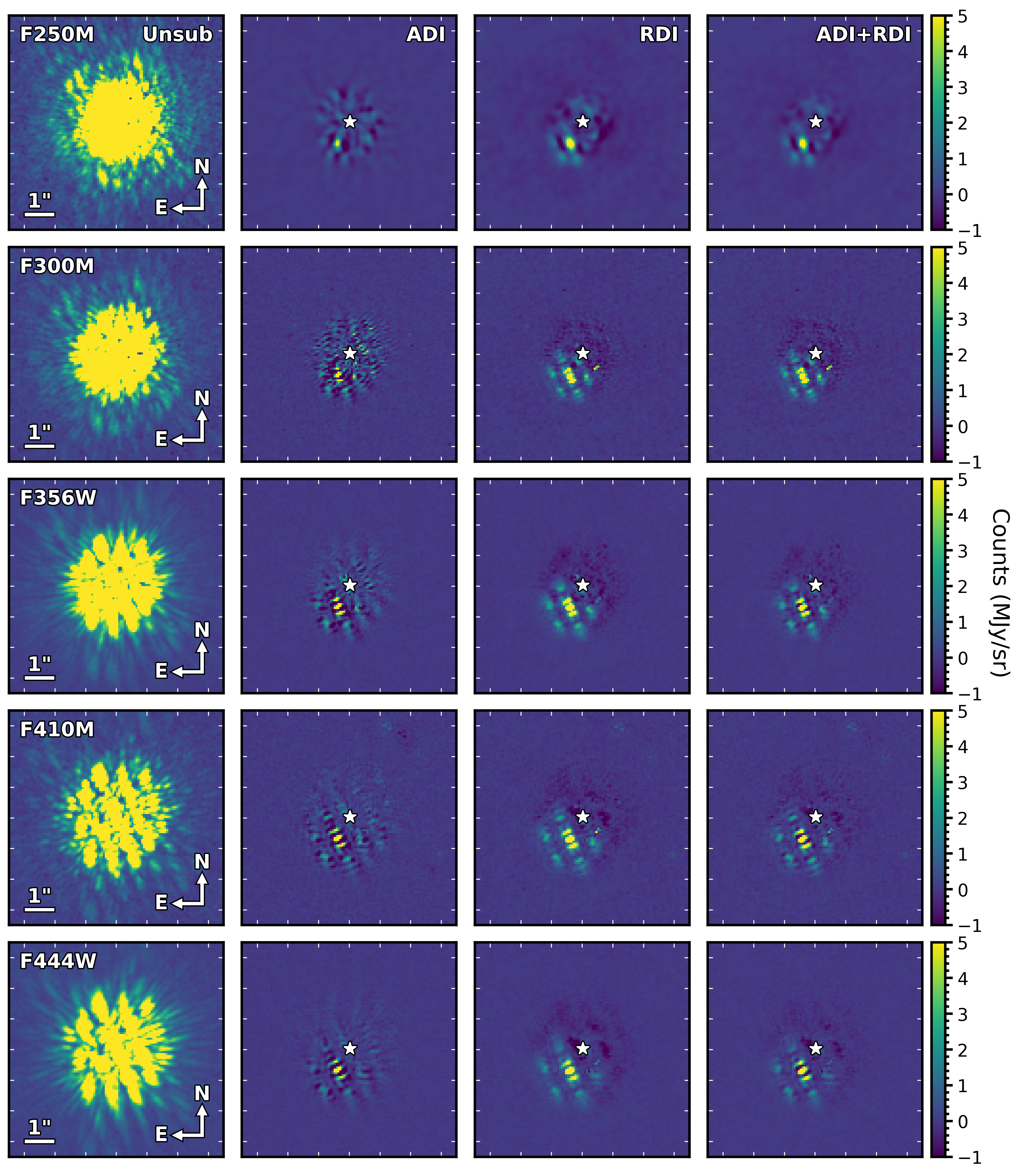}
    \caption{As in \fref{fig:two_images}, except for the NIRCam F250M, F300M, F356W, F410M, and F444W filters. Here we show subtractions using the maximum number of PCA modes for ADI, RDI, and ADI+RDI, respectively. The F250M subtracted images have been smoothed as described in Section \ref{sec:obs}.}
    \label{fig:nircam_images}
\end{figure*}
\begin{figure*}[h]
    \centering
    \includegraphics[width=0.97\textwidth]{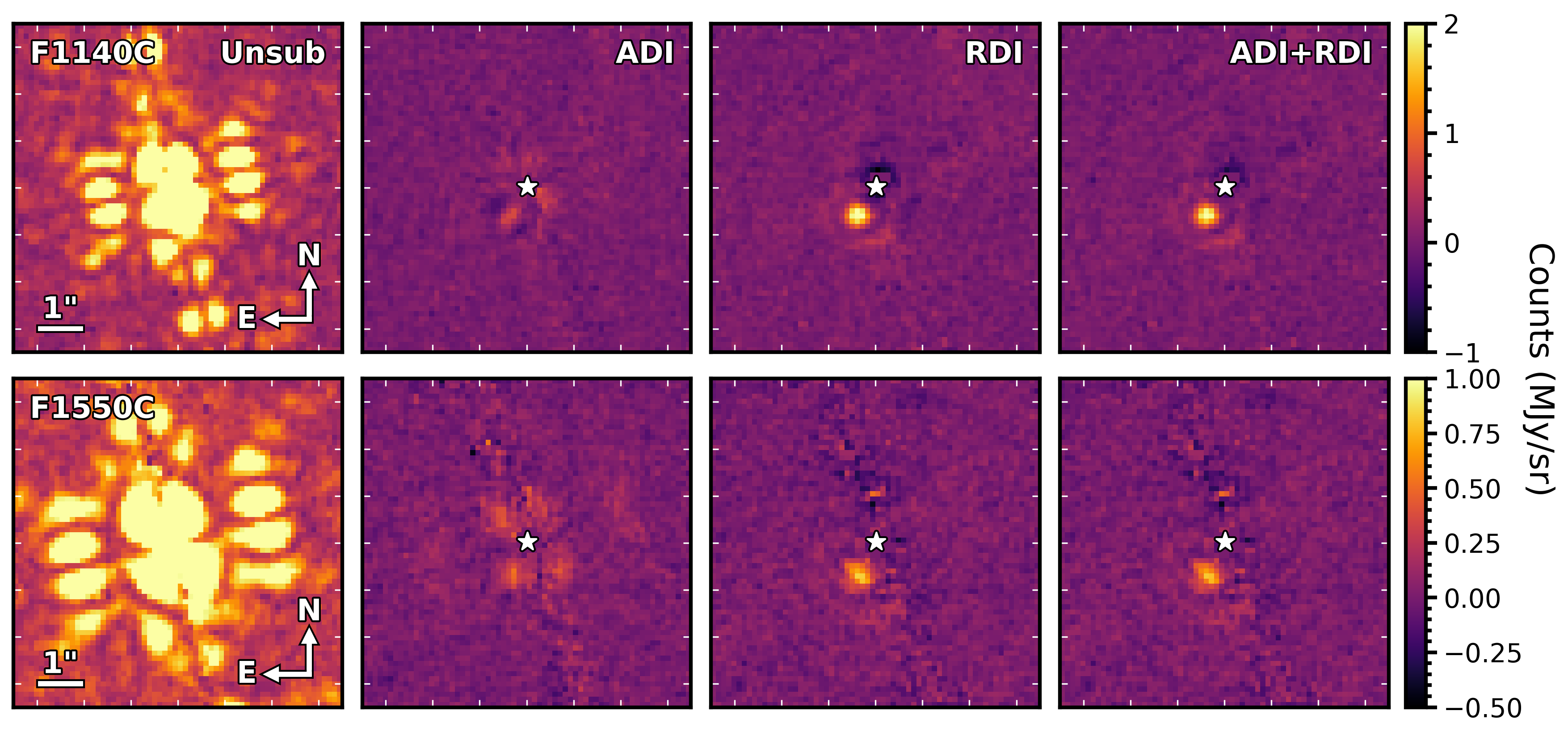}
    \caption{As in \fref{fig:two_images}, except for the MIRI F1140C and F1550C filters. Here we show subtractions using the maximum number of PCA modes for ADI, RDI, and ADI+RDI, respectively. To aid visual clarity, the subtracted F1550C images are shown with a peak image intensity five times smaller than the unsubtracted image.}
    \label{fig:miri_images}
\end{figure*}

\clearpage
\section{Contrast Performance}\label{app:ccurves}
\begin{figure*}[b]
    \centering
    \includegraphics[width=\textwidth]{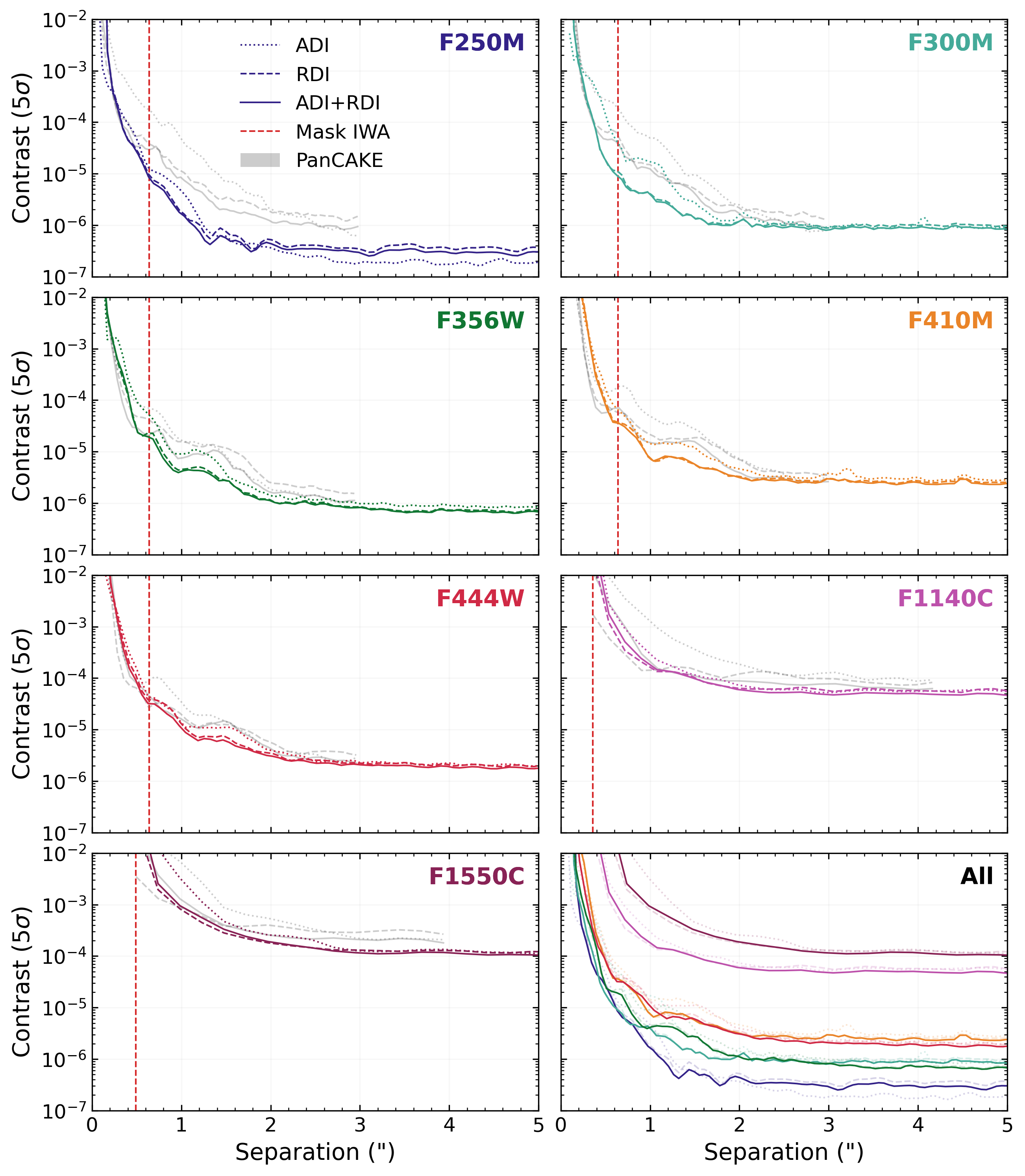}
    \caption{As in \fref{fig:contrast_curves}, but for all used filters. We also plot the equivalent predicted contrast curves for these observations from PanCAKE (grey lines) following \citep{Cart21b}. In every filter \textit{JWST} is exceeding its predicted performance.}
    \label{fig:all_contrasts}
\end{figure*}
\clearpage
\section{PSF Fitting}\label{app:psffits}
\vspace{-5mm}
\begin{figure*}[h]
    \centering
    \includegraphics[width=0.76\textwidth]{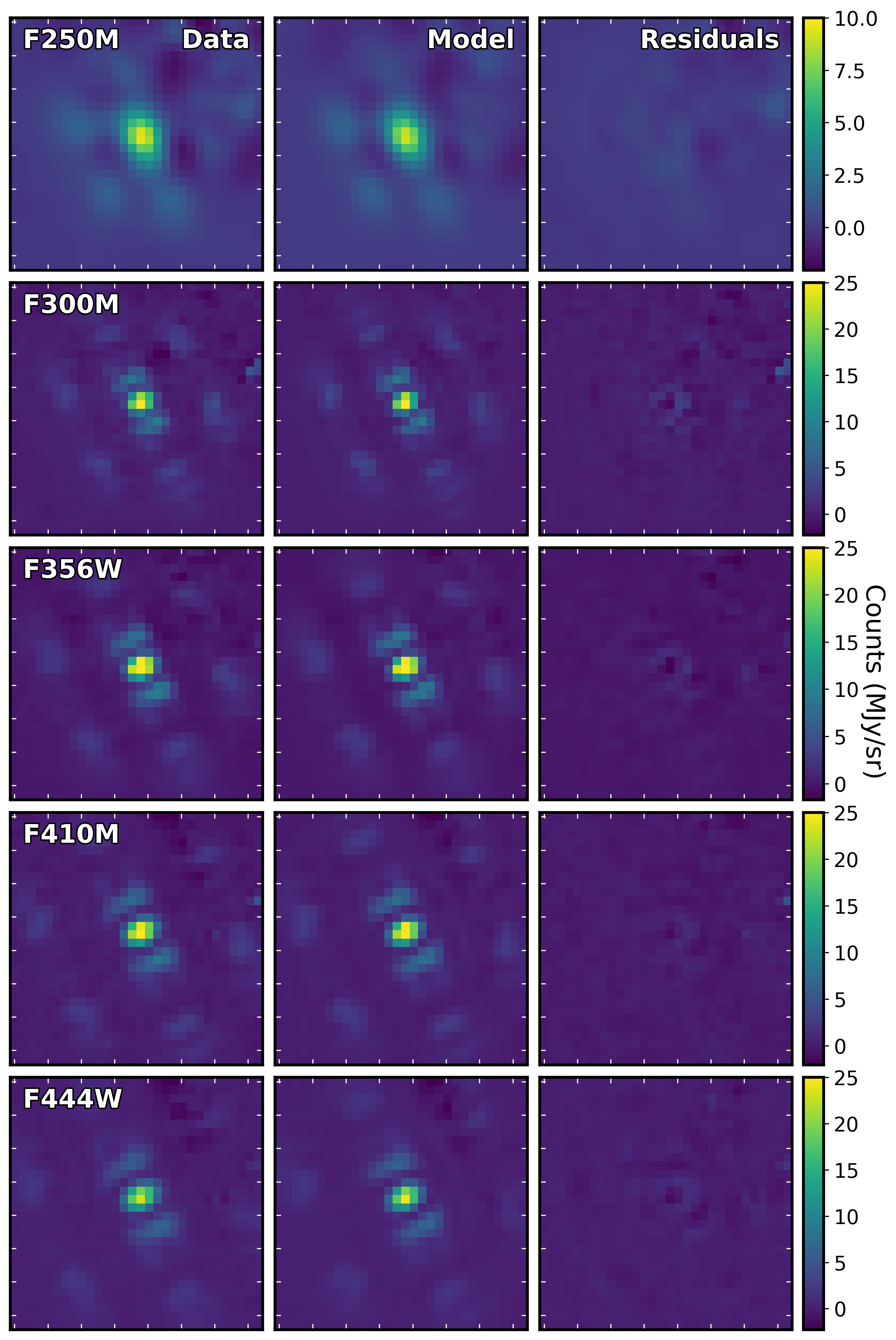}
    \caption{The data (left column), model PSF (middle column), and residuals (right column) for the \sklip{} PSF fitting of the NIRCam observations of \target{}. Pixel counts are in MJy/sr and are indicated by the colour bar on the right hand side, images are oriented with North upwards and North through East counterclockwise, and the image size is 30$\times$30 pixels (1.9$\times$1.9\arcsec).} 
    \label{fig:nircam_imfit}
\end{figure*}
\begin{figure*}[h]
    \centering
    \includegraphics[width=0.76\textwidth]{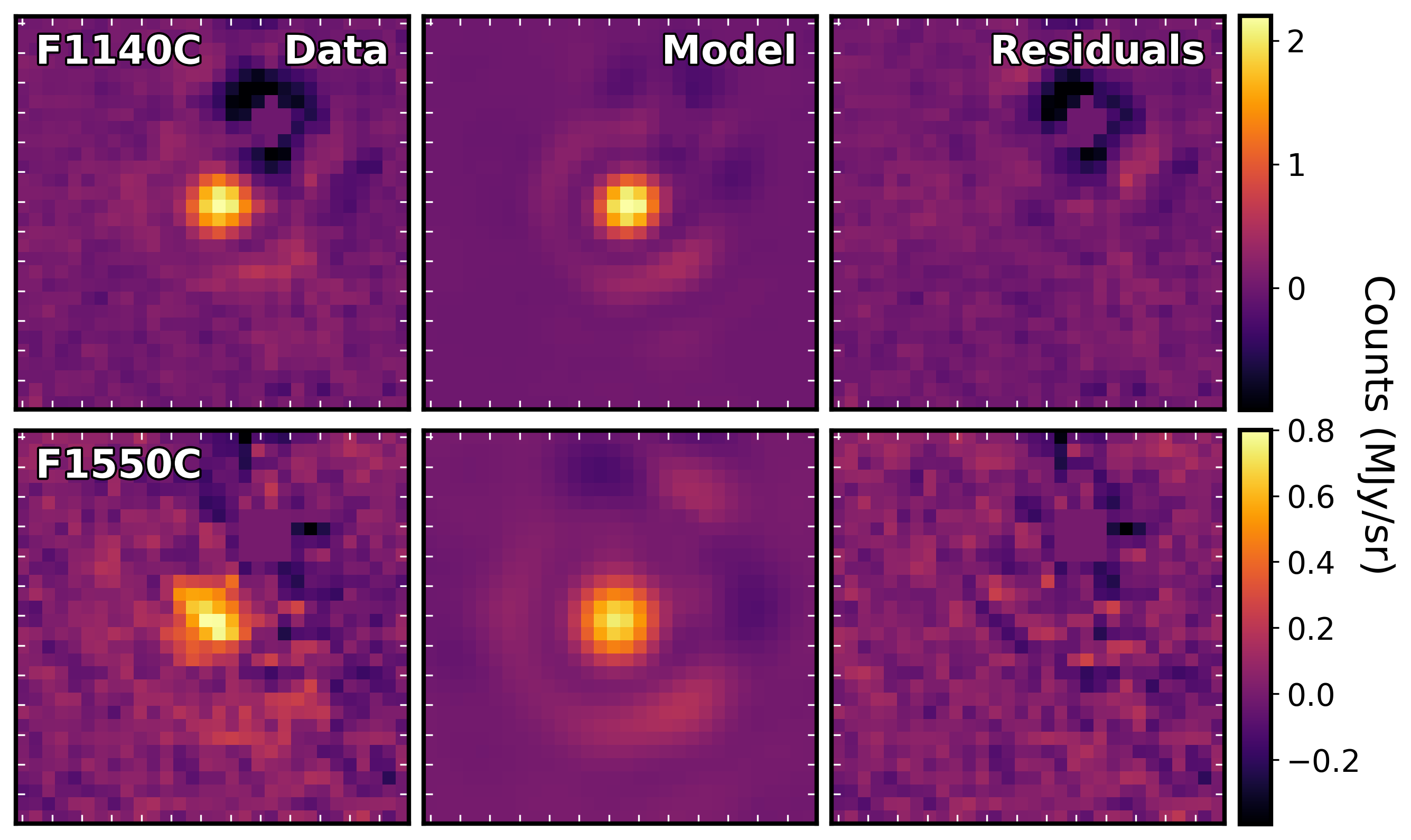}
    \caption{As in \fref{fig:nircam_imfit}, but for the MIRI observations. The image size is 30$\times$30 pixels (3.3$\times$3.3\arcsec).}
    \label{fig:miri_imfit}
\end{figure*}
\clearpage
\section{Complementary photometric measurements of HIP\,65426\,b}\label{app:phottable}
\vspace{-5mm}
\startlongtable
\begin{deluxetable*}{cccccc}
    \renewcommand\arraystretch{1}
    \tablecaption{Additional photometric measurements of HIP\,65426\,b considered in this work.}
    %\centering\setlength{\tabcolsep}{6pt}
    %\begin{tabular}{llllll}
    %\multicolumn{6}{c}{{\bfseries \tablename\ \thetable{}.} -- }\\
    % \hline
    % \hline
    % \vspace{1mm}
    % Wavelength & Bandwidth & Instrument & Band & Flux & Ref.\\
    % ($\mu$m) & ($\mu$m) & & & (W m$^{-2} \mu$m$^{-1}$) & \\
    % \hline %\vspace{-4mm} \\
    % \endfirsthead
    % \multicolumn{6}{c}{{\bfseries \tablename\ \thetable{}.} -- Continued.}\\
    % \hline
    % \hline
    % Wavelength & Bandwidth & Instrument & Band & Flux & Ref.\\
    % ($\mu$m) & ($\mu$m) & & & (W m$^{-2} \mu$m$^{-1}$) & \\
    % \hline %\vspace{-4mm} \\
    % \endhead
    % \hline
    % \endfoot
    % \endlastfoot
    \tablehead{\colhead{Wavelength ($\mu$m)} & \colhead{Bandwidth ($\mu$m)} & \colhead{Instrument} & \colhead{Band} & \colhead{Flux (W m$^{-2} \mu$m$^{-1}$)} & \colhead{Ref.}}
    \startdata
    1.002 & 0.011 & SPHERE/IFS & YJ & $(2.434 \pm 0.569) \times 10^{-17}$ & 1\\
    1.011 & 0.011 & SPHERE/IFS & YJ & $(3.155 \pm 0.670) \times 10^{-17}$ & 1\\
    1.021 & 0.011 & SPHERE/IFS & YJ & $(3.564 \pm 0.575) \times 10^{-17}$ & 1\\
    1.030 & 0.011 & SPHERE/IFS & YJ & $(3.199 \pm 0.428) \times 10^{-17}$ & 1\\
    1.040 & 0.011 & SPHERE/IFS & YJ & $(4.095 \pm 0.482) \times 10^{-17}$ & 1\\
    1.050 & 0.011 & SPHERE/IFS & YJ & $(3.829 \pm 0.435) \times 10^{-17}$ & 1\\
    1.060 & 0.011 & SPHERE/IFS & YJ & $(4.154 \pm 0.489) \times 10^{-17}$ & 1\\
    1.070 & 0.011 & SPHERE/IFS & YJ & $(4.456 \pm 0.521) \times 10^{-17}$ & 1\\
    1.081 & 0.011 & SPHERE/IFS & YJ & $(4.679 \pm 0.432) \times 10^{-17}$ & 1\\
    1.091 & 0.011 & SPHERE/IFS & YJ & $(5.143 \pm 0.483) \times 10^{-17}$ & 1\\
    1.102 & 0.011 & SPHERE/IFS & YJ & $(4.746 \pm 0.543) \times 10^{-17}$ & 1\\
    1.112 & 0.011 & SPHERE/IFS & YJ & $(5.127 \pm 0.576) \times 10^{-17}$ & 1\\
    1.123 & 0.011 & SPHERE/IFS & YJ & $(4.775 \pm 0.641) \times 10^{-17}$ & 1\\
    1.133 & 0.011 & SPHERE/IFS & YJ & $(5.228 \pm 0.682) \times 10^{-17}$ & 1\\
    1.144 & 0.011 & SPHERE/IFS & YJ & $(4.982 \pm 0.609) \times 10^{-17}$ & 1\\
    1.154 & 0.011 & SPHERE/IFS & YJ & $(4.821 \pm 0.519) \times 10^{-17}$ & 1\\
    1.165 & 0.011 & SPHERE/IFS & YJ & $(5.324 \pm 0.486) \times 10^{-17}$ & 1\\
    1.175 & 0.011 & SPHERE/IFS & YJ & $(4.739 \pm 0.453) \times 10^{-17}$ & 1\\
    1.186 & 0.011 & SPHERE/IFS & YJ & $(5.709 \pm 0.477) \times 10^{-17}$ & 1\\
    1.196 & 0.011 & SPHERE/IFS & YJ & $(5.291 \pm 0.415) \times 10^{-17}$ & 1\\
    1.206 & 0.011 & SPHERE/IFS & YJ & $(6.155 \pm 0.471) \times 10^{-17}$ & 1\\
    1.217 & 0.011 & SPHERE/IFS & YJ & $(6.586 \pm 0.490) \times 10^{-17}$ & 1\\
    1.227 & 0.011 & SPHERE/IFS & YJ & $(6.431 \pm 0.500) \times 10^{-17}$ & 1\\
    1.237 & 0.011 & SPHERE/IFS & YJ & $(6.203 \pm 0.483) \times 10^{-17}$ & 1\\
    1.247 & 0.011 & SPHERE/IFS & YJ & $(6.409 \pm 0.477) \times 10^{-17}$ & 1\\
    1.257 & 0.011 & SPHERE/IFS & YJ & $(7.100 \pm 0.483) \times 10^{-17}$ & 1\\
    1.266 & 0.011 & SPHERE/IFS & YJ & $(7.289 \pm 0.502) \times 10^{-17}$ & 1\\
    1.276 & 0.011 & SPHERE/IFS & YJ & $(7.344 \pm 0.504) \times 10^{-17}$ & 1\\
    1.285 & 0.011 & SPHERE/IFS & YJ & $(7.315 \pm 0.499) \times 10^{-17}$ & 1\\
    1.294 & 0.011 & SPHERE/IFS & YJ & $(8.215 \pm 0.543) \times 10^{-17}$ & 1\\
    1.303 & 0.011 & SPHERE/IFS & YJ & $(8.701 \pm 0.543) \times 10^{-17}$ & 1\\
    1.312 & 0.011 & SPHERE/IFS & YJ & $(8.731 \pm 0.580) \times 10^{-17}$ & 1\\
    1.321 & 0.011 & SPHERE/IFS & YJ & $(7.450 \pm 0.594) \times 10^{-17}$ & 1\\
    1.329 & 0.011 & SPHERE/IFS & YJ & $(7.073 \pm 0.583) \times 10^{-17}$ & 1\\
    0.987 & 0.019 & SPHERE/IFS & YJH & $(1.556 \pm 0.312) \times 10^{-17}$ & 1\\
    1.002 & 0.019 & SPHERE/IFS & YJH & $(1.791 \pm 0.355) \times 10^{-17}$ & 1\\
    1.018 & 0.019 & SPHERE/IFS & YJH & $(2.855 \pm 0.510) \times 10^{-17}$ & 1\\
    1.034 & 0.019 & SPHERE/IFS & YJH & $(2.615 \pm 0.305) \times 10^{-17}$ & 1\\
    1.051 & 0.019 & SPHERE/IFS & YJH & $(3.394 \pm 0.442) \times 10^{-17}$ & 1\\
    1.068 & 0.019 & SPHERE/IFS & YJH & $(4.239 \pm 0.456) \times 10^{-17}$ & 1\\
    1.086 & 0.019 & SPHERE/IFS & YJH & $(3.472 \pm 0.407) \times 10^{-17}$ & 1\\
    1.104 & 0.019 & SPHERE/IFS & YJH & $(3.802 \pm 0.415) \times 10^{-17}$ & 1\\
    1.122 & 0.019 & SPHERE/IFS & YJH & $(4.403 \pm 0.504) \times 10^{-17}$ & 1\\
    1.140 & 0.019 & SPHERE/IFS & YJH & $(3.906 \pm 0.397) \times 10^{-17}$ & 1\\
    1.159 & 0.019 & SPHERE/IFS & YJH & $(4.132 \pm 0.422) \times 10^{-17}$ & 1\\
    1.178 & 0.019 & SPHERE/IFS & YJH & $(4.641 \pm 0.452) \times 10^{-17}$ & 1\\
    1.197 & 0.019 & SPHERE/IFS & YJH & $(5.368 \pm 0.544) \times 10^{-17}$ & 1\\
    1.216 & 0.019 & SPHERE/IFS & YJH & $(6.356 \pm 0.573) \times 10^{-17}$ & 1\\
    1.235 & 0.019 & SPHERE/IFS & YJH & $(6.763 \pm 0.602) \times 10^{-17}$ & 1\\
    1.255 & 0.019 & SPHERE/IFS & YJH & $(7.107 \pm 0.622) \times 10^{-17}$ & 1\\
    1.274 & 0.019 & SPHERE/IFS & YJH & $(7.228 \pm 0.630) \times 10^{-17}$ & 1\\
    1.294 & 0.019 & SPHERE/IFS & YJH & $(7.599 \pm 0.668) \times 10^{-17}$ & 1\\
    1.314 & 0.019 & SPHERE/IFS & YJH & $(7.296 \pm 0.648) \times 10^{-17}$ & 1\\
    1.333 & 0.019 & SPHERE/IFS & YJH & $(6.046 \pm 0.577) \times 10^{-17}$ & 1\\
    1.353 & 0.019 & SPHERE/IFS & YJH & $(5.400 \pm 0.649) \times 10^{-17}$ & 1\\
    1.372 & 0.019 & SPHERE/IFS & YJH & $(5.523 \pm 0.872) \times 10^{-17}$ & 1\\
    1.391 & 0.019 & SPHERE/IFS & YJH & $(5.875 \pm 0.729) \times 10^{-17}$ & 1\\
    1.411 & 0.019 & SPHERE/IFS & YJH & $(5.285 \pm 0.550) \times 10^{-17}$ & 1\\
    1.430 & 0.019 & SPHERE/IFS & YJH & $(4.516 \pm 0.460) \times 10^{-17}$ & 1\\
    1.449 & 0.019 & SPHERE/IFS & YJH & $(4.768 \pm 0.439) \times 10^{-17}$ & 1\\
    1.467 & 0.019 & SPHERE/IFS & YJH & $(5.374 \pm 0.484) \times 10^{-17}$ & 1\\
    1.486 & 0.019 & SPHERE/IFS & YJH & $(5.888 \pm 0.506) \times 10^{-17}$ & 1\\
    1.504 & 0.019 & SPHERE/IFS & YJH & $(5.927 \pm 0.498) \times 10^{-17}$ & 1\\
    1.522 & 0.019 & SPHERE/IFS & YJH & $(6.259 \pm 0.519) \times 10^{-17}$ & 1\\
    1.539 & 0.019 & SPHERE/IFS & YJH & $(6.778 \pm 0.561) \times 10^{-17}$ & 1\\
    1.556 & 0.019 & SPHERE/IFS & YJH & $(7.318 \pm 0.605) \times 10^{-17}$ & 1\\
    1.573 & 0.019 & SPHERE/IFS & YJH & $(7.560 \pm 0.624) \times 10^{-17}$ & 1\\
    1.589 & 0.019 & SPHERE/IFS & YJH & $(8.037 \pm 0.665) \times 10^{-17}$ & 1\\
    1.605 & 0.019 & SPHERE/IFS & YJH & $(8.497 \pm 0.706) \times 10^{-17}$ & 1\\
    1.621 & 0.019 & SPHERE/IFS & YJH & $(8.579 \pm 0.715) \times 10^{-17}$ & 1\\
    1.636 & 0.019 & SPHERE/IFS & YJH & $(9.011 \pm 0.762) \times 10^{-17}$ & 1\\
    1.593 & 0.055 & SPHERE/IRDIS & H2 & $(8.569 \pm 0.383) \times 10^{-17}$ & 1\\
    1.667 & 0.056 & SPHERE/IRDIS & H3 & $(10.129 \pm 0.564) \times 10^{-17}$ & 1\\
    2.110 & 0.102 & SPHERE/IRDIS & K1 & $(7.500 \pm 0.600) \times 10^{-17}$ & 1\\
    2.251 & 0.109 & SPHERE/IRDIS & K2 & $(7.100 \pm 0.600) \times 10^{-17}$ & 1\\
    3.800 & 0.620 & NACO & L’ & $(4.010 \pm 0.542) \times 10^{-17}$ & 2, 3\\
    4.051 & 0.020 & NACO & NB4.05 & $(3.220 \pm 0.780) \times 10^{-17}$ & 3\\
    4.780 & 0.590 & NACO & M’ & $(2.549 \pm 0.820) \times 10^{-17}$ & 2, 3 \\
    \enddata
    %\end{tabular}
    \tablerefs{(1) \citet{Chau17}; (2) \citet{Chee19}; (3) \citet{Stol20}.}
    \tablecomments{For the $L'$ and $M'$-band photometry, we considered the average of the measurements reported in \citet{Chee19} and \citet{Stol20}, but kept the largest error bars.}% \\ {\bf References}: $^1$\citet{Chau17}, $^2$\citet{Chee19}, $^3$\citet{Stol20}.}
    %\caption{{\bf References}: \\
             % {\bf Notes}: For the $L'$ and $M'$-band photometry, we considered the average of the measurements reported in \citet{Chee19} and \citet{Stol20}, but kept the largest error bars}}}
    % }
    \label{tab:addphot}
\end{deluxetable*}
\end{document}